\documentclass[american,aps,pra,reprint,superscriptaddress,longbibliography]{revtex4-2}
\usepackage[T1]{fontenc}
\usepackage[latin9]{inputenc}
\usepackage{color}
\usepackage{babel}
\usepackage{amsthm}
\usepackage{amsmath}
\usepackage{amssymb}
\usepackage{graphicx}
\usepackage{dsfont}
\usepackage{physics}
\usepackage{todonotes}
\usepackage[unicode=true,pdfusetitle, bookmarks=true,bookmarksnumbered=false,bookmarksopen=false,  breaklinks=false,pdfborder={0 0 0},backref=false,colorlinks=false] {hyperref}
\hypersetup{ colorlinks,linkcolor=myurlcolor,citecolor=myurlcolor,urlcolor=myurlcolor}

\makeatletter
\@ifundefined{textcolor}{}
{%
	\definecolor{BLACK}{gray}{0}
	\definecolor{WHITE}{gray}{1}
	\definecolor{RED}{rgb}{1,0,0}
	\definecolor{GREEN}{rgb}{0,1,0}
	\definecolor{BLUE}{rgb}{0,0,1}
	\definecolor{CYAN}{cmyk}{1,0,0,0}
	\definecolor{MAGENTA}{cmyk}{0,1,0,0}
	\definecolor{YELLOW}{cmyk}{0,0,1,0}
}
\theoremstyle{plain}

\theoremstyle{plain}

\ifx\proof\undefined
\newenvironment{proof}[1][\protect\proofname]{\par
	\normalfont\topsep6\p@\@plus6\p@\relax
	\trivlist
	\itemindent\parindent
	\item[\hskip\labelsep
	\scshape
	#1]\ignorespaces
}{%
	\endtrivlist\@endpefalse
}
\providecommand{\proofname}{Proof}
\fi
\theoremstyle{plain}

\providecommand{\lemmaname}{Lemma}
\providecommand{\definitionname}{Definition}
\providecommand{\propositionname}{Proposition}

\usepackage{babel}
\usepackage{txfonts}
\usepackage{colortbl}\definecolor{myurlcolor}{rgb}{0,0,0.7}
\def\proj#1{| #1 \rangle\!\langle #1 |}

\newcommand{\haH}

\definecolor{orange}{RGB}{255,127,0}
\newcommand{\mnb}[1]{{\color{blue} #1}}

\begin{document}
	\title{
		Ancilla-Assisted Protection of Information: Application to Atom-Cavity Systems}
	\author{Rajeev Gangwar}
	\affiliation{Department of Physical Sciences, Indian Institute of Science Education and Research (IISER), Mohali, Punjab 140306, India}
	\author{Mohit Lal Bera}
	\affiliation{ICFO -- Institut de Ci\`encies Fot\`oniques, The Barcelona Institute of Science and Technology, ES-08860 Castelldefels, Spain}
	\author{G. P. Teja }
	\affiliation{Department of Physical Sciences, Indian Institute of Science Education and Research (IISER), Mohali, Punjab 140306, India}
	\author{Sandeep K.~Goyal}
	\affiliation{Department of Physical Sciences, Indian Institute of Science Education and Research (IISER), Mohali, Punjab 140306, India}
	\author{Manabendra Nath Bera}
	\email{mnbera@gmail.com}
	\affiliation{Department of Physical Sciences, Indian Institute of Science Education and Research (IISER), Mohali, Punjab 140306, India}
	
	\begin{abstract}
		One of the major obstacles faced by quantum-enabled technology is the environmental noise that causes decoherence in the quantum system, thereby destroying much of its quantum aspects and introducing errors while the system undergoes quantum operations and processing. A number of techniques have been invented to mitigate the environmental effects, and many of these techniques are specific to the environment and the quantum tasks at hand. Here, we propose a protocol that makes arbitrary environments effectively noise-free or transparent using an ancilla, which, in particular, is well suited to protect information stored in atoms. The ancilla, which is the photons, is allowed to undergo restricted but a wide class of noisy operations. The protocol transfers the information of the system onto the decoherence-free subspace and later retrieves it back to the system. Consequently, it enables full protection of quantum information and entanglement in the atomic system from decoherence. We propose experimental schemes to implement this protocol on atomic systems in an optical cavity.
	\end{abstract}

	\maketitle
	
	\onecolumngrid
	
	\section{Introduction}
	Quantum-enabled technologies involve well-controlled systems to store information and processing, such as atomic and solid state devices, and systems that are suitable for networking and communication, such as photonic systems. For information processing, it is often necessary to protect coherence in the state of a quantum system for a long enough duration. However, a quantum system cannot be fully isolated from its environment, and the latter induces decoherence in the system. These effects thereby destroy much of its quantum aspects, e.g., quantum coherence, entanglement, etc., and introduces uncontrolled errors in information processing~\cite{Unruh95, Breuer02, Zurek03, nielsen2010}. Taking over environmental noise is thus one of the major challenges in quantum-enabled technologies today. 
	
	Some of the well-known techniques to eliminate environmental effects include dynamical decoupling~\cite{Seth_1999, Khodjasteh05, Liu13}, weak measurements, Zeno effect \cite{Kondo_2016} and coherent feedback control~\cite{Bluhm10} to suppress the decoherence, quantum error-correction~\cite{Gottesman97, Aoki09, Yao12, Terhal15, nielsen2010}, error-mitigating methods \cite{Temme17, Endo18, McArdle19, Kandala19, Kwiat00, Blume08}, and use of decoherence-free subspace \cite{Zanardi97, Lidar98, Bacon00,Wu_2021} for quantum computation and simulations. 
	
	Dynamical decoupling is an open-loop quantum control technique. It is implemented by a periodic sequence of instantaneous control pulses \cite{Seth_1999} and achieves decoherence suppression without increasing the Hilbert space dimension. Still, it cannot be applied to non-Markovian processes \cite{Addis_2015, Robin_Hillier2021} (e.g., decoherence due to amplitude damping). In the decoherence-free subspace approach, quantum information is encoded in a particular quantum state that does not experience a specific type of decoherence \cite{Lidar1998, Paul_2000}. Both the quantum error correction and the decoherence-free subspace-based schemes use the Hilbert space dimension larger than that of the system dimension. The quantum Zeno effect is also exploited to protect a quantum system \mnb{\cite{Kondo_2016}}. There are other interesting methods to protect a quantum state and entanglement distribution using weak measurement and their reversals \cite{He13, Doustimotlagh_2014, Kim:09, Lim:14, kim2012protecting, Sun2010, Zong2014, Yao2012, Man2012, Starling2013, Royer1994, Korotkov2006, Sun2009} and decoherence suppression via quantum measurement reversal \cite{Lee:11}. In these schemes, the quantum state is firstly transferred to more robust states by a weak measurement to resist decoherence. After that, another weak measurement is performed that reverses the state back to the original state. Due to the failure rates of the weak measurements, however, these schemes have limited success probabilities.
	Weak force sensing is also a technique to protect the quantum state, and this is based on coherent quantum noise cancellation in a non-linear hybrid optomechanical system \cite{singh2023}. The optomechanical cavity contains a movable mechanical mirror, a fixed semitransparent mirror, an ensemble of ultra-cold atoms, and an optical parametric amplifier (OPA). Using the coherent quantum noise cancellation (CQNC) process, one can eliminate the back action noise at all frequencies. Also, by tuning the OPA parameters, one can suppress the quantum shot noise at lower frequencies than the resonant frequency.
There are also techniques to suppress noise in an atomic system using a field in a squeezed coherent state \cite{gelman2010}. The interaction of a quantized electromagnetic field with a medium of three-level $\Lambda$ atoms has been studied adequately in the last several years \cite{Dantan2004, Dantan2005, Barberis2007}. The interaction of a quantized electromagnetic field in a squeezed coherent state with a three-level $\Lambda$ atom is studied numerically by the quantum Monte Carlo method and analytically by the Heisenberg-Langevin method in the regime of electromagnetically induced transparency (EIT) \cite{J.P_1998}.

Here we introduce a generic protocol that is particularly suitable for storing quantum information in an atomic system placed inside an optical cavity. It makes arbitrary noisy environment noise-free for the atom using externally supplied photons. The protocol enables full protection of quantum information and entanglement from decoherence and environmental noise. We also propose an experimentally realizable scheme for that.
	
\begin{figure}
		\includegraphics[width=0.9 \columnwidth]{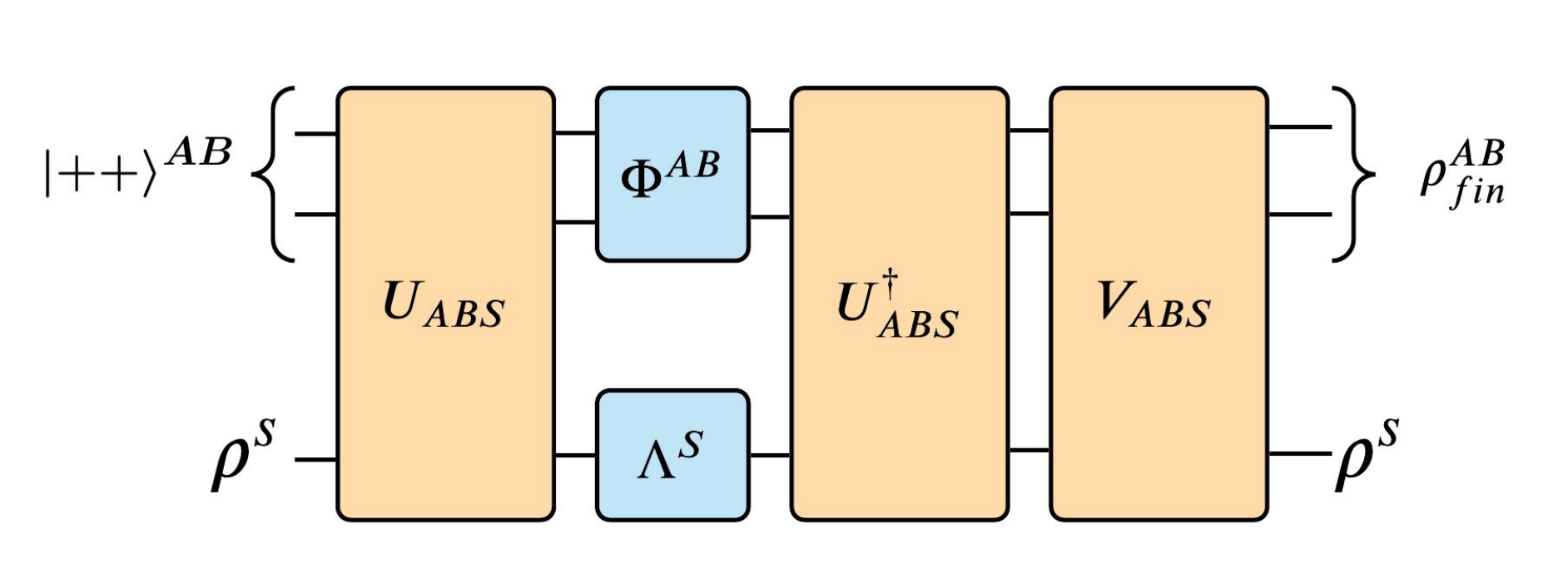}
		\caption{{\bf Quantum circuit for QCT:}  The protocol uses two ancilla $A$ and $B$ in the initial state $\ket{++}^{AB}$. In the pre-processing phase, the non-local unitary $U_{ABS}$ is applied on the system $S$ and the ancilla $AB$, which couples the three qubits before they are exposed to the environments. Due to the environment, the system $S$ undergoes an arbitrary noisy operation $\Lambda^S$, while the ancilla $AB$ may be exposed to a wide class of noise given by the map $\Phi^{AB}$ in Eq.~\eqref{eq:5}. In the post-processing phase, we undo the operation $U_{ABS}$, i.e., apply $U_{ABS}^\dagger$, followed by the non-local unitary operation $V_{ABS}$, which neutralizes the effect of environmental noise on $S$. 
			\label{fig:CircuitQCT}}
	\end{figure}

To mitigate noise from a system, the protocol requires an ancilla.  For instance, the system may be an atom, and the ancilla may be two photons, as we shall consider later. The protocol exploits non-local evolutions on the system and ancilla. In the process, the information stored in the system is transferred onto the decoherence-free subspace \cite{Lidar98, Bacon00} of the ancilla before the system is exposed to environmental noise and retrieved back to the system at the end. The action of an arbitrary channel $\Lambda^S$ on a state $\rho^S$ of a $d$-dimensional system (qudit) $S$ can be expressed as
	\begin{align} \label{eq:GenChan}
		\Lambda^S(\rho^S)= \sum_m F_m^S \ \rho^S \ F_m^{S\dag}, 
	\end{align}
	where $F_m^S$s are the Kraus operators satisfying $\sum_m F_m^{S\dag} F_m^S=\mathds{1}$. Here, $\mathds{1}$ is the $d \times d$ identity matrix. The protocol implements the transformation of the operation on the system part from arbitrary noisy operation $\Lambda$ to identity operation as, 
	\begin{align*}
		\Lambda^S \to \mathds{I}^S.
	\end{align*}
The protocol achieves a transformation in the system's operation, transforming it from an arbitrary noisy operation $\Lambda^S$ to the identity operation $\mathds{I}^S$, where $\mathds{I}^S$ signifies the identity channel acting on the system. The goal of this paper is to introduce a protocol implementing such a transformation on the system's operations. The protocol relies on two ancilla and unitary operations on the composite (see Fig. \ref{fig:CircuitQCT}). We also introduce an experimentally implementable model based on an atom-cavity system. We discuss all these in the following sections.

\section{Protection of information against noise  \label{sec:QCT}}
	Let us denote two qudits $A$ and $B$ as ancilla. The system is in an arbitrary state $\rho^S$. A quantum circuit scheme to realize the protocol is given in figure~\ref{fig:CircuitQCT}. 
 \\
	{\bf The noise mitigation protocol} --  Without loss of generality, we consider a qubit ($d=2$) system, which is exposed to arbitrary qubit noisy channels. The protocol can be extended to an arbitrary $d$-dimensional atom (see Appendix). For $d=2$, the Kraus operators in Eq.~\eqref{eq:GenChan} can be expressed in terms of superposition in the orthonormal Hilbert-Schmidt bases, as $F_m^S=\sum_{i=0, x, y, z}c_{mi} \ \sigma^S_i$, where $c_{mi} \in \mathds{C}$, $\sigma_0=\mathds{1}$, and $\sigma^S_x, \sigma^S_y$, and $\sigma^S_z$ are the Pauli $x, y$, and $z$ matrices, respectively. Throughout the text we denote $\ket{0}$ and $\ket{1}$ as the eigenstates of $\sigma^S_z$ operator. The step-by-step description of the protocol is given below (see Appendix for more details). \\
	
	\begin{enumerate}
		\item[Step 1:] The ancilla $AB$ is initiated in the state $\ket{++}$ where $\ket{+}^{A/B}=(\ket{0}^{A/B}+ \ket{1}^{A/B})/\sqrt{2}$.
		\item[Step 2:] The composite $ABS$ is evolved by a global unitary operation $U_{ABS}$, given by
		\begin{align}\label{eq:Umax1}
			U_{ABS}= & \ \proj{00}^{AB} \otimes  \mathds{1}^S + \proj{01}^{AB} \otimes \sigma_z^S \nonumber \\
			& + \proj{10}^{AB} \otimes \sigma_x^S - i \proj{11}^{AB} \otimes \sigma^S_y.
		\end{align}
		\item[Step 3:] The system $S$ undergoes an arbitrary (possibly unknown) qubit channel $\Lambda^S$, as a result of its interaction with its environment. Similarly, the ancilla may be allowed to experience environmental noise resulting in a specific class of channels  $\Phi^{AB}$. The form of Kraus operators of the channels $\Phi^{AB}=\sum_\mu E^{AB}_\mu \rho_{AB} (E^{AB}_\mu)^\dag$ are given by
		\begin{align}
			E^{AB}_\mu = q_{\mu 0} \mathds{1} \otimes \mathds{1} + 	q_{\mu 1} \sigma_x \otimes \mathds{1} + q_{\mu 2} \sigma_y \otimes \sigma_x + q_{\mu 3} \sigma_z \otimes \sigma_x, \label{eq:5}
		\end{align}  
		where $q_{\mu i} \in \mathds{C}$ and $\sum_\mu E_\mu^{AB \dagger} E_\mu^{AB}=\mathds{1}^{AB}$. The form the Kraus operators depend on the idea of decoherence-free subspace which we shall discuss later and obvious example of the operation $\Phi^{AB}$ is the identity operation. 
		
		\item[ Step 4:] The composite $ABS$ is evolved with the unitary $U_{ABS}^\dag$ followed by another unitary $V_{ABS}$ given by
		\begin{align}\label{eq:Umax2}
			V_{ABS}= & \ \proj{++}^{AB} \otimes \mathds{1}^S+ \proj{+-}^{AB} \otimes \sigma_x^S \nonumber \\
			& +i \ \proj{--}^{AB} \otimes \sigma_y^S + \proj{-+}^{AB} \otimes \sigma_z^S.
		\end{align}
		
	\end{enumerate}
	
	On the level of channel, the Steps 1-4 implement a transformation that leads to 
	\begin{align}\label{eq:Is}
		\Phi^{AB} \otimes \Lambda^S \rightarrow \Psi^{AB} \otimes \mathds{I}^{S},  
	\end{align}
	for an arbitrary noisy channel $\Lambda^S$ on $S$. Note that, in place of applying the unitary $V_{ABS}$ in Step 4, we may also perform a non-unitary operation with the Kraus operators $\{ \proj{++}^{AB} \otimes \mathds{1}^S, \ \proj{+-}^{AB} \otimes \sigma_x^S, \ \proj{--}^{AB} \otimes \sigma_y^S,  \ \proj{-+}^{AB} \otimes \sigma_z^S\}$ to make the channel $\Lambda^S$ transparent. \\

{\bf Role of decoherence-free subspace} -- The protocol transfers the information of the initial system state to the ancilla after Step 2, i.e., before the system is exposed to arbitrary local noise. And the information transferred to the ancilla does not alter as it is encoded onto the decoherence-free subspace \cite{Lidar98, Bacon00} corresponding to the noise given by the Kraus operators \eqref{eq:5}. In addition, arbitrary local noise on the system does not degrade the information stored in ancilla either. Step 4 only retrieves the information from the ancilla to the system. Thus, the system recovers its initial state at the end of the protocol. 

For a given set of channels $\Phi$ on some system with the Kraus operators $E_\mu$ satisfying $\sum_\mu E_\mu^\dag E_\mu=\mathds{1}$, we define a decoherence-free subspace $\mathcal{H}_d$ if 
\begin{align}
	\rho = \Phi(\rho)= \sum_\mu E_\mu \rho E_\mu^\dag,
\end{align}
where $\rho$ is state that live in $\mathcal{H}_d$. Note, this alternatively requires that $[\rho, E_\mu]=0, \ \forall \mu$. This notion of the decoherence-free subspace is similar to the one introduced in \cite{Lidar98, Bacon00} except that it is defined for a quantum channel without having an explicit semi-group structure.

Now, we discuss how the protocol transfers the information about the initial state of the system $S$ to the ancilla $AB$. The global state after Step 2 (see above) is  
\begin{align}
	\rho_2^{ABS}=U_{ABS}\rho_1^{ABS}U_{ABS}^\dagger, \nonumber
\end{align}
where $	\rho^{ABS}_1=\proj{++}^{AB}\otimes\rho^S$, and $\rho^S$ is an arbitrary state of $S$. Now the reduced density matrix of $AB$ is given by 
\begin{align}
	\rho_{AB}=\frac{1}{4}\Bigg[\Big(\mathds{1}\otimes\mathds{1}\Big)+r_z\Big(\mathds{1}\otimes\sigma_x\Big)-r_y\Big(\sigma_x\otimes\sigma_y\Big)+r_x\Big(\sigma_x\otimes\sigma_z\Big)\Bigg],	
\end{align} 
where $r_i = \tr(\rho^S \sigma_i)$ with $i=x,y,z$. Note, all the information about $\rho^S$ is transferred to $AB$ in terms of $\{r_x, r_y, r_z\}$ and no information remains in $S$ as its reduced state becomes maximally mixed after the evolution. 

In Step 3, the ancilla $AB$ is exposed to the noise channels $\Phi^{AB}$ with the Kraus operators $E_\mu^{AB}=\sum_{ij=0}^1q_{\mu ij}\sigma_z^i\sigma_x^j\otimes\sigma_x^{2-i}$.  But the interesting fact is that the reduced state $\rho^{AB}$ does not alter by this evolution, i.e., 
\begin{align}
	\rho^{AB}=\Phi^{AB}(\rho^{AB}) = \sum_\mu E_\mu^{AB} \rho^{AB} E_\mu^{AB \dag}.	
\end{align}
It implies that the information about $S$, in terms of  $\{r_x, r_y, r_z\}$, is stored in a subspace of $AB$ that is decoherence-free for the class of noisy channels given by $\Phi^{AB}$. Further, an arbitrary local noisy evolution ($\Lambda^S$) on the system $S$ cannot destroy or degrade the information about $S$ being stored in $AB$. After Step 4, the information about $S$ is retrieved, which is how it is protected from arbitrary local noises.  

Clearly, arbitrary noisy channels on the system ($S$) can be made transparent using the above protocol if the ancilla ($AB$) is restricted to experiencing a particular class of noises. Not only that, but for a multipartite system, it can protect entanglement while the subsystems are exposed to local environments. Beyond these restricted noises on ancilla, the protocol does not always protect quantum information in the system in general. However, there are realistic cases where the ancilla ($AB$) does not experience the same noise as the system does. For instance, the atoms and photons undergo different noisy evolutions when they are exposed to the `same' environment. Below we explore such a situation where an atom in a cavity is considered as the system undergoing arbitrary noisy evolution, and photons act as the ancilla.
	
	\section{Protecting atoms in optical cavities}
	
\begin{figure}
		\includegraphics[width=1 \columnwidth]{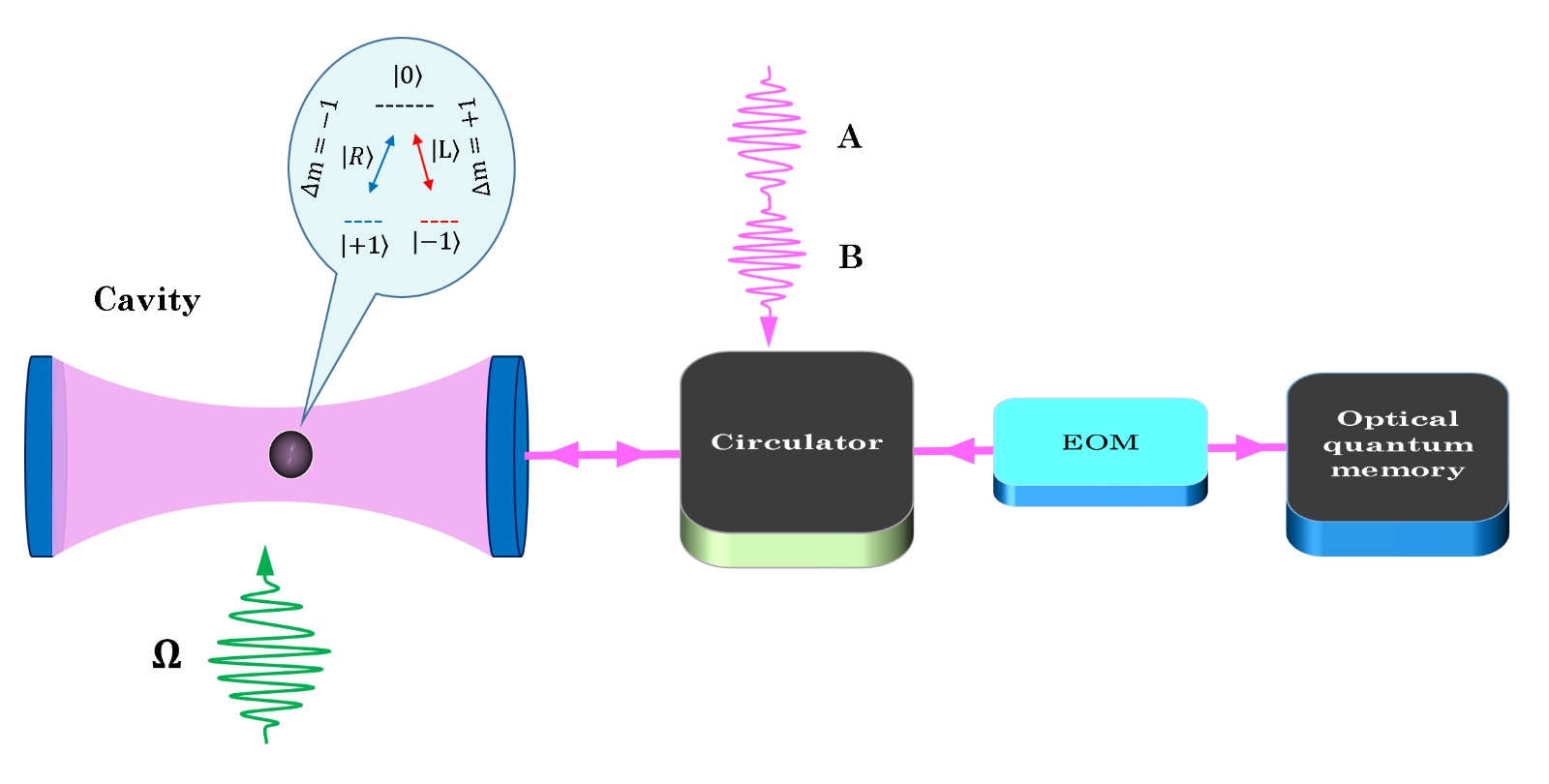}
		\caption{{\bf Protecting quantum information in an atomic system:} The atomic state space is spanned by the low-energy states $\ket{\pm 1}$ and the excited state $\ket{0}$. The ancilla consists of two single-photons ($A$ and $B$) in circularly polarization basis $\{\ket{R},\ket{L}\}$. The photons are initiated in the states $(\ket{R}+\ket{L})/\sqrt{2}$. In order to implement $U_{ABS}$ operation, we make two photons $A$ and $B$ interact with the atom $S$ inside the cavity. Let the two photons be $\tau$ time apart where the photon $B$ comes first. The interaction of the photon $B$ with the atom results in the $C_z^{BS}$ operation. Then applying Hadamard operation on the atom using STIRAP technique with the help of a classical laser pulses ($\Omega$) followed by the interaction of the photon $A$ yields $C_x^{AS}$ operation. Both operations together result in $U_{ABS}$. Afterward, the two photons can be stored in an optical quantum memory \cite{wang19, guo19, hosseini11} to be used subsequently to implement $U^\dagger_{ABS}$ and $V_{ABS}$ operations. Circulators \cite{hacker2016} are used to direct the photons towards the cavity and optical quantum memory. The electro-optic-modulator (EOM) \cite{luo2015, amin2018} is used to perform Hadamard operation on the photons in order to convert $U_{ABS}$ into $V_{ABS}$.
		}
		\label{fig:QCT_cavity}
	\end{figure}

	In quantum information processing tasks, it is often necessary to protect coherence in the state of a quantum system for a long enough duration. But, the inevitable interaction with the environment makes the quantum coherence in a state to decay. The environmental effects can be nullified using the noise mitigation protocol introduced above hence protecting the state and the quantum information indefinitely. To demonstrate that, we present an experimental scheme in Figure~\ref{fig:QCT_cavity}, where the state of an atom is protected for an indefinite time.
	
	In this scheme, we consider a three-level atom with a two-fold degenerate ground state and an excited state trapped inside an optical cavity. The system qubit $(S)$ consists of the state space spanned by the low-energy states $\ket{\pm 1}$ of the atom, whereas the ancilla qubits ($A$ and $B$) consist of two single-photons.
	
	To implement the operation $U_{ABS}$ we first notice that this operator can be decomposed as a product of two controlled two-qubit operators as follows:
	\begin{align}
		U_{ABS}=\big(\mathds{1}_B\otimes C_x^{AS}\big) \big( \mathds{1}_A\otimes C_z^{BS}\big),
	\end{align}
	where $C_x^{AS}=\dyad{0}^A \otimes \mathds{1}^S+ \dyad{1}^A \otimes \sigma_x^S$ and $C_z^{BS}=\dyad{0}^B \otimes \mathds{1}^S +  \dyad{1}^B \otimes \sigma_z^S$ are respectively the control-NOT and control-Phase gates acting on one photon and the atom.
	
	The control-Phase and the control-NOT operations between a photon and the atom can be performed using the technique given in Ref.~\cite{reiserer2014quantum,hacker2016,kim2013quantum,duan2004scalable}. For this technique to work, the three-level atom should have the  $\Lambda$-transition, where the transition is allowed only between $\ket{\pm 1} \leftrightarrow \ket{0}$ and forbidden between $\ket{+1}\leftrightarrow\ket{-1}$ levels. Due to the conservation of angular momentum, the right-circular polarized light interacts with $\ket{+1} \leftrightarrow \ket{0}$ transition and the left-circular with $\ket{-1} \leftrightarrow \ket{0}$.
	
	For an atom trapped inside a cavity in the strong coupling regime (Purcell regime), in the steady state limit and at resonance, the relation between the input $a_{\text{in}}$ and output $a_{\text{out}}$ light modes of the cavity can be written as
	\begin{align}
		a_{\text{out}} = \frac{-\kappa\gamma + 4g^2}{\kappa\gamma + 4g^2}a_{\text{in}},
	\end{align}
	where  $\kappa$ is the decay rate of the cavity, $g$ characterizes the coupling strength between the cavity and the atom and $\gamma$ is the atomic decay rate such that  $\kappa \gg g \gg \gamma$.
	Therefore, if $g^2 \gg \kappa\gamma$ then $a_{\text{out}} \sim a_{\text{in}}$ whereas if $g^2 \ll \kappa\gamma$ then $a_{\text{out}} \sim -a_{\text{in}}$.

	If we choose the right-circular polarization as the normal mode of the optical cavity, then  the transition $\ket{-1}\leftrightarrow \ket{0}$ is always decoupled, i.e, $g = 0$.
	Therefore, if the atom is prepared in the state $\ket{-1}$ then  it will reflect the photon with a $\pi$-phase, irrespective of its polarization. However, if the atom is in the state $\ket{+1}$ then the photon will experience a $\sigma_z$ operation on its polarization states, i.e., $\ket{L} \to -\ket{L}$ and $\ket{R} \to \ket{R}$. The general transformation can be written as
	\begin{equation}\label{gz}
		\ket{L} \otimes \ket{\pm1}  \to -\ket{L} \otimes \ket{\pm1},\quad
		\ket{R} \otimes \ket{\pm1}  \to  \pm\ket{R} \otimes \ket{\pm1},
	\end{equation}
which is exactly $-C_z$ operation. The $C_z$ operation can be converted into the $C_x$ operation by applying the Hadamard operation on the target qubit. It can be done efficiently on the atomic system by the optimized-STIRAP techniques where the excited state $\ket{0}$ is adiabatically eliminated \cite{baksic2016, chen2010}. Optimized-STIRAP has duration around $\sim 100 ns$ \cite{zhou2017} and the typical errors are less than $10^{-4}$ \cite{Vitanov_2017,wu2022}. To perform controlled operations, we use a single atom ($\Lambda$-system) trapped in a cavity. This $\Lambda$-system is made up of three levels: $\ket{0}$, $\ket{+1}$, and $\ket{-1}$. When transitioning from $\ket{0}$ to $\ket{\pm 1}$, the coherence time is typically around $ (\gamma/2)^{-1}$, approximately $\sim 1\mu s $. As discussed above, our qubit ($\ket{+1}$, $\ket{-1}$) is created by eliminating the excited state $\ket{0}$, making the influence of decoherence (from $\gamma$) negligible on the overall operation time. To conduct controlled operations, we need light pulses with widths much larger than $\kappa/g^2$. Therefore, boosting the coupling strength $g$ can speed up the execution of controlled operations. However, in modern experiments with atoms in cavities, pulses lasting $\sim 1\mu s $ to $\sim 2\mu s $ are commonly used \cite{hacker2019}. As for $U_{ABS}$, it involves two reflections separated by an atomic rotation, and with current atom-cavity settings, the estimated time is around $\sim 5\mu s $. Additionally, atomic rotations are performed using STIRAP, a process that takes about $\sim 100 ns$ \cite{zhou2017}. It's crucial to highlight that the extended time required for controlled operations is mainly due to the input pulse width. Nevertheless, this duration can be significantly shortened by increasing the coupling strength, bringing the total time to around a few $\sim 100 ns$ \cite{zhou2017}. Controlled operations between atoms and a cavity exhibit low fidelities. For instance, a CNOT operation is executed with an $86\%$ fidelity \cite{hacker2016}. This reduction in fidelity primarily arises from frequency fluctuations and scattering losses associated with the cavity mirrors rather than cavity decay rates. Hence, by tuning the parameters $(g, \kappa, \gamma)$, an atom-cavity system can perform unitary operations with minimal losses, primarily due to photon loss via scattering from mirrors.

	Therefore, the unitary $U_{ABS}$ can be implemented by sequentially interacting the two photons with the atom-cavity system. Similarly, we can implement the $V_{ABS}$ operator between the atom and the photons by introducing $U_{ABS}$ along with the Hadamard operation on the polarization states of the photons using HWP and the atomic states using STIRAP technique. 
 
	As mentioned earlier, one of the advantages of photonic systems is that they experience less environmental noise than for atomic systems. In fact, the major source of photonic noise is associated with the loss of the photon itself.  However, recent experiments demonstrate that, in atom-cavity systems, the photon loss can be restricted to less than $5\%$ \cite{Reiserer14, Hacker16, Welte18}. This, in turn, implies there is a $0.05$ probability with which a photon is lost. Now we shall study the impact of photon loss in the atom-cavity system discussed above. There are several possibilities. For instance, photon-$A$ may get lost in different stages of the protocol, i.e., in step-2 or in step-4, and similarly for photon-$B$. Further, both the photons $A$ and $B$ may get lost in different stages. 
	
	Consider the photon-$A$ is lost before step 2 takes place. Because of that, the effect of environmental noise on the atom cannot be eliminated. Say, the atom undergoes an evolution given by the channel $\Phi^S_2$. This channel will be different for different environmental noises the atom is exposed to. For the cases where the photon is being lost with probability $p$, the overall evolution of the atom is $\mathcal{E}^S=(1-p) \mathds{I}^S + p \Phi^S_2$, where $\mathds{I}^S$ represents the identity operation when there is no photon loss (see Eq.~\eqref{eq:Is}). Now, it can be easily seen that the error incurred in the noise mitigation due to the photon loss for any given initial state of the atom is 
	\begin{align}
		|| \mathds{I}^S - \mathcal{E}^S||_\Diamond = p || \mathds{I}^S - \Phi^S_2||_\Diamond \leq p,
	\end{align}    	
	where $|| \cdot ||_\Diamond$ is the diamond norm \cite{Benenti_2010}. Similarly, for the case where a photon is lost in step 4 with a probability $p$, the error in noise mitigation is bounded from above by $p$. This is also true when the photon-$B$ is lost in either step 2 or 4. In the case where both the photon $A$ and $B$ are lost with probabilities $p$ and $q$, respectively, the overall error in noise mitigation is upper bounded by $pq$ irrespective of the steps in which the photons are lost.  
	
	\section{Discussion}
	Unlike the conventional noise-mitigating protocols that are, in general, specific to the system,  and the nature of the environment, our protocol can be applicable to an arbitrary channel on any system. Although, it requires the ancilla to undergo a restricted class of noisy channels. If one allows arbitrary noise on the ancilla, the proposed protocol can only mitigate noise in some specific class noises on the system and its initial states. The methods to improve coherence time using dynamical decoupling and feedback control are particular to the nature of noise in the system. While these protect local coherence to an extent, they are not suitable for preserving non-local coherence, such as quantum entanglement. In contrast, our protocol can protect both local and global coherence for an indefinite time, irrespective of the nature of local noises on the systems. There are techniques to protect quantum entanglement by creating it in a decoherence-free subspace. However, these are specific to a small set of entangled states and do not work for arbitrary states. Instead, in our protocol, a decoherence-free subspace of the ancilla is utilized to eliminate arbitrary local noise in the system. Another technique to mitigate the environmental effects is the SWAP protocol, where one can transfer the noise effects from the system to the ancilla by swapping their states. However, this requires the ancilla to be noise-free at all times. On the contrary, our protocol works even if the ancilla is allowed to interact with a large but restricted class of noisy environments. 
	
	Our protocol is particularly suitable for various practical situations where different kinds of systems (or degrees of freedom) experience different noises while in the same physical environment. For instance, a photon experiences different noises than an atom. We exploit this advantage to construct an experimental proposal where an atom in a cavity is made noise-free with the use of photons. 
	
	In summary, we have introduced a protocol that makes an arbitrary quantum channel transparent for an arbitrary dimensional quantum system. Specifically, we have given the implementation scheme to protect quantum information in an atomic system inside a cavity. The protocol may open new avenues to protect quantum information and correlations against environmental noise. \\
	
	{\bf Acknowledgments} -- R.G. thanks the Council of Scientific and Industrial Research (CSIR), Government of India, for financial support through a fellowship (File No. 09/947(0233)/2019-EMR-I). S.K.G. acknowledges the financial support from Inter-disciplinary Cyber Physical Systems(ICPS) program of the Department of Science and Technology, India, (Grant No. DST/ICPS/QuST/Theme-1/2019/12). M.N.B. gratefully acknowledges financial supports from SERB-DST (CRG/2019/002199), Government of India. M.L.B. thankfully acknowledges supports from ERC AdG NOQIA, Agencia Estatal de Investigaci\'on (``Severo Ochoa'' Center of Excellence CEX2019-000910-S, Plan National FIDEUA PID2019-106901GB-I00/10.13039/501100011033, FPI), Fundaci\'o Privada Cellex, Fundaci\'o Mir-Puig, and from Generalitat de Catalunya (AGAUR Grant No. 2017 SGR 1341, CERCA program, QuantumCAT\_U16-011424, co-funded by ERDF Operational Program of Catalonia 2014-2020), MINECO-EU QUANTERA MAQS (funded by State Research Agency (AEI) PCI2019-111828-2/10.13039/501100011033), EU Horizon 2020 FET-OPEN OPTOLogic (Grant No 899794), and the National Science Centre, Poland-Symfonia Grant No. 2016/20/W/ST4/00314.

	{\bf Competing interests} - The authors declare no competing
	interests.
	
	{\bf Data and materials availability} - Data sharing not applicable
	to this article, as no datasets were generated or analyzed
	during the current study.


\begin{thebibliography}{66}%
\makeatletter
\providecommand \@ifxundefined [1]{%
 \@ifx{#1\undefined}
}%
\providecommand \@ifnum [1]{%
 \ifnum #1\expandafter \@firstoftwo
 \else \expandafter \@secondoftwo
 \fi
}%
\providecommand \@ifx [1]{%
 \ifx #1\expandafter \@firstoftwo
 \else \expandafter \@secondoftwo
 \fi
}%
\providecommand \natexlab [1]{#1}%
\providecommand \enquote  [1]{``#1''}%
\providecommand \bibnamefont  [1]{#1}%
\providecommand \bibfnamefont [1]{#1}%
\providecommand \citenamefont [1]{#1}%
\providecommand \href@noop [0]{\@secondoftwo}%
\providecommand \href [0]{\begingroup \@sanitize@url \@href}%
\providecommand \@href[1]{\@@startlink{#1}\@@href}%
\providecommand \@@href[1]{\endgroup#1\@@endlink}%
\providecommand \@sanitize@url [0]{\catcode `\\12\catcode `\$12\catcode
  `\&12\catcode `\#12\catcode `\^12\catcode `\_12\catcode `\%12\relax}%
\providecommand \@@startlink[1]{}%
\providecommand \@@endlink[0]{}%
\providecommand \url  [0]{\begingroup\@sanitize@url \@url }%
\providecommand \@url [1]{\endgroup\@href {#1}{\urlprefix }}%
\providecommand \urlprefix  [0]{URL }%
\providecommand \Eprint [0]{\href }%
\providecommand \doibase [0]{https://doi.org/}%
\providecommand \selectlanguage [0]{\@gobble}%
\providecommand \bibinfo  [0]{\@secondoftwo}%
\providecommand \bibfield  [0]{\@secondoftwo}%
\providecommand \translation [1]{[#1]}%
\providecommand \BibitemOpen [0]{}%
\providecommand \bibitemStop [0]{}%
\providecommand \bibitemNoStop [0]{.\EOS\space}%
\providecommand \EOS [0]{\spacefactor3000\relax}%
\providecommand \BibitemShut  [1]{\csname bibitem#1\endcsname}%
\let\auto@bib@innerbib\@empty
\bibitem [{\citenamefont {Unruh}(1995)}]{Unruh95}%
  \BibitemOpen
  \bibfield  {author} {\bibinfo {author} {\bibfnamefont {W.~G.}\ \bibnamefont
  {Unruh}},\ }\bibfield  {title} {\bibinfo {title} {Maintaining coherence in
  quantum computers},\ }\href {https://doi.org/10.1103/PhysRevA.51.992}
  {\bibfield  {journal} {\bibinfo  {journal} {Phys. Rev. A}\ }\textbf {\bibinfo
  {volume} {51}},\ \bibinfo {pages} {992} (\bibinfo {year} {1995})}\BibitemShut
  {NoStop}%
\bibitem [{\citenamefont {Breuer}\ and\ \citenamefont
  {Petruccione}(2002)}]{Breuer02}%
  \BibitemOpen
  \bibfield  {author} {\bibinfo {author} {\bibfnamefont {H.~P.}\ \bibnamefont
  {Breuer}}\ and\ \bibinfo {author} {\bibfnamefont {F.}~\bibnamefont
  {Petruccione}},\ }\href
  {https://doi.org/10.1093/acprof:oso/9780199213900.001.0001} {\emph {\bibinfo
  {title} {The theory of open quantum systems}}}\ (\bibinfo  {publisher}
  {Oxford University Press},\ \bibinfo {address} {Great Clarendon Street},\
  \bibinfo {year} {2002})\BibitemShut {NoStop}%
\bibitem [{\citenamefont {Zurek}(2003)}]{Zurek03}%
  \BibitemOpen
  \bibfield  {author} {\bibinfo {author} {\bibfnamefont {W.~H.}\ \bibnamefont
  {Zurek}},\ }\bibfield  {title} {\bibinfo {title} {Decoherence, einselection,
  and the quantum origins of the classical},\ }\href
  {https://doi.org/10.1103/RevModPhys.75.715} {\bibfield  {journal} {\bibinfo
  {journal} {Rev. Mod. Phys.}\ }\textbf {\bibinfo {volume} {75}},\ \bibinfo
  {pages} {715} (\bibinfo {year} {2003})}\BibitemShut {NoStop}%
\bibitem [{\citenamefont {Nielsen}\ and\ \citenamefont
  {Chuang}(2010)}]{nielsen2010}%
  \BibitemOpen
  \bibfield  {author} {\bibinfo {author} {\bibfnamefont {M.~A.}\ \bibnamefont
  {Nielsen}}\ and\ \bibinfo {author} {\bibfnamefont {I.~L.}\ \bibnamefont
  {Chuang}},\ }\href {https://doi.org/10.1017/CBO9780511976667} {\emph
  {\bibinfo {title} {Quantum Computation and Quantum Information: 10th
  Anniversary Edition}}}\ (\bibinfo  {publisher} {Cambridge University Press},\
  \bibinfo {year} {2010})\BibitemShut {NoStop}%
\bibitem [{\citenamefont {Viola}\ \emph {et~al.}(1999)\citenamefont {Viola},
  \citenamefont {Knill},\ and\ \citenamefont {Lloyd}}]{Seth_1999}%
  \BibitemOpen
  \bibfield  {author} {\bibinfo {author} {\bibfnamefont {L.}~\bibnamefont
  {Viola}}, \bibinfo {author} {\bibfnamefont {E.}~\bibnamefont {Knill}},\ and\
  \bibinfo {author} {\bibfnamefont {S.}~\bibnamefont {Lloyd}},\ }\bibfield
  {title} {\bibinfo {title} {Dynamical decoupling of open quantum systems},\
  }\href {https://doi.org/10.1103/PhysRevLett.82.2417} {\bibfield  {journal}
  {\bibinfo  {journal} {Phys. Rev. Lett.}\ }\textbf {\bibinfo {volume} {82}},\
  \bibinfo {pages} {2417} (\bibinfo {year} {1999})}\BibitemShut {NoStop}%
\bibitem [{\citenamefont {Khodjasteh}\ and\ \citenamefont
  {Lidar}(2005)}]{Khodjasteh05}%
  \BibitemOpen
  \bibfield  {author} {\bibinfo {author} {\bibfnamefont {K.}~\bibnamefont
  {Khodjasteh}}\ and\ \bibinfo {author} {\bibfnamefont {D.~A.}\ \bibnamefont
  {Lidar}},\ }\bibfield  {title} {\bibinfo {title} {Fault-tolerant quantum
  dynamical decoupling},\ }\href
  {https://doi.org/10.1103/PhysRevLett.95.180501} {\bibfield  {journal}
  {\bibinfo  {journal} {Phys. Rev. Lett.}\ }\textbf {\bibinfo {volume} {95}},\
  \bibinfo {pages} {180501} (\bibinfo {year} {2005})}\BibitemShut {NoStop}%
\bibitem [{\citenamefont {Liu}\ \emph {et~al.}(2013)\citenamefont {Liu},
  \citenamefont {Po}, \citenamefont {Du}, \citenamefont {Liu},\ and\
  \citenamefont {Pan}}]{Liu13}%
  \BibitemOpen
  \bibfield  {author} {\bibinfo {author} {\bibfnamefont {G.-Q.}\ \bibnamefont
  {Liu}}, \bibinfo {author} {\bibfnamefont {H.~C.}\ \bibnamefont {Po}},
  \bibinfo {author} {\bibfnamefont {J.}~\bibnamefont {Du}}, \bibinfo {author}
  {\bibfnamefont {R.-B.}\ \bibnamefont {Liu}},\ and\ \bibinfo {author}
  {\bibfnamefont {X.-Y.}\ \bibnamefont {Pan}},\ }\bibfield  {title} {\bibinfo
  {title} {Noise-resilient quantum evolution steered by dynamical decoupling},\
  }\href {https://doi.org/10.1038/ncomms3254} {\bibfield  {journal} {\bibinfo
  {journal} {Nat. Commun.}\ }\textbf {\bibinfo {volume} {4}},\ \bibinfo {pages}
  {2254} (\bibinfo {year} {2013})}\BibitemShut {NoStop}%
\bibitem [{\citenamefont {Kondo}\ \emph {et~al.}(2016)\citenamefont {Kondo},
  \citenamefont {Matsuzaki}, \citenamefont {Matsushima},\ and\ \citenamefont
  {Filgueiras}}]{Kondo_2016}%
  \BibitemOpen
  \bibfield  {author} {\bibinfo {author} {\bibfnamefont {Y.}~\bibnamefont
  {Kondo}}, \bibinfo {author} {\bibfnamefont {Y.}~\bibnamefont {Matsuzaki}},
  \bibinfo {author} {\bibfnamefont {K.}~\bibnamefont {Matsushima}},\ and\
  \bibinfo {author} {\bibfnamefont {J.~G.}\ \bibnamefont {Filgueiras}},\
  }\bibfield  {title} {\bibinfo {title} {Using the quantum zeno effect for
  suppression of decoherence},\ }\href
  {https://doi.org/10.1088/1367-2630/18/1/013033} {\bibfield  {journal}
  {\bibinfo  {journal} {New J. Phys.}\ }\textbf {\bibinfo {volume} {18}},\
  \bibinfo {pages} {013033} (\bibinfo {year} {2016})}\BibitemShut {NoStop}%
\bibitem [{\citenamefont {Bluhm}\ \emph {et~al.}(2010)\citenamefont {Bluhm},
  \citenamefont {Foletti}, \citenamefont {Mahalu}, \citenamefont {Umansky},\
  and\ \citenamefont {Yacoby}}]{Bluhm10}%
  \BibitemOpen
  \bibfield  {author} {\bibinfo {author} {\bibfnamefont {H.}~\bibnamefont
  {Bluhm}}, \bibinfo {author} {\bibfnamefont {S.}~\bibnamefont {Foletti}},
  \bibinfo {author} {\bibfnamefont {D.}~\bibnamefont {Mahalu}}, \bibinfo
  {author} {\bibfnamefont {V.}~\bibnamefont {Umansky}},\ and\ \bibinfo {author}
  {\bibfnamefont {A.}~\bibnamefont {Yacoby}},\ }\bibfield  {title} {\bibinfo
  {title} {Enhancing the coherence of a spin qubit by operating it as a
  feedback loop that controls its nuclear spin bath},\ }\href
  {https://doi.org/10.1103/PhysRevLett.105.216803} {\bibfield  {journal}
  {\bibinfo  {journal} {Phys. Rev. Lett.}\ }\textbf {\bibinfo {volume} {105}},\
  \bibinfo {pages} {216803} (\bibinfo {year} {2010})}\BibitemShut {NoStop}%
\bibitem [{\citenamefont {Gottesman}(1997)}]{Gottesman97}%
  \BibitemOpen
  \bibfield  {author} {\bibinfo {author} {\bibfnamefont {D.}~\bibnamefont
  {Gottesman}},\ }\emph {\bibinfo {title} {Stabilizer Codes and Quantum Error
  Correction}},\ \href {https://doi.org/doi:10.7907/rzr7-dt72} {Ph.D. thesis}
  (\bibinfo {year} {1997})\BibitemShut {NoStop}%
\bibitem [{\citenamefont {Aoki}\ \emph {et~al.}(2009)\citenamefont {Aoki},
  \citenamefont {Takahashi}, \citenamefont {Kajiya}, \citenamefont {Yoshikawa},
  \citenamefont {Braunstein}, \citenamefont {van Loock},\ and\ \citenamefont
  {Furusawa}}]{Aoki09}%
  \BibitemOpen
  \bibfield  {author} {\bibinfo {author} {\bibfnamefont {T.}~\bibnamefont
  {Aoki}}, \bibinfo {author} {\bibfnamefont {G.}~\bibnamefont {Takahashi}},
  \bibinfo {author} {\bibfnamefont {T.}~\bibnamefont {Kajiya}}, \bibinfo
  {author} {\bibfnamefont {J.-i.}\ \bibnamefont {Yoshikawa}}, \bibinfo {author}
  {\bibfnamefont {S.~L.}\ \bibnamefont {Braunstein}}, \bibinfo {author}
  {\bibfnamefont {P.}~\bibnamefont {van Loock}},\ and\ \bibinfo {author}
  {\bibfnamefont {A.}~\bibnamefont {Furusawa}},\ }\bibfield  {title} {\bibinfo
  {title} {Quantum error correction beyond qubits},\ }\href
  {https://doi.org/10.1038/nphys1309} {\bibfield  {journal} {\bibinfo
  {journal} {Nat. Phys.}\ }\textbf {\bibinfo {volume} {5}},\ \bibinfo {pages}
  {541} (\bibinfo {year} {2009})}\BibitemShut {NoStop}%
\bibitem [{\citenamefont {Yao}\ \emph {et~al.}(2012{\natexlab{a}})\citenamefont
  {Yao}, \citenamefont {Wang}, \citenamefont {Chen}, \citenamefont {Gao},
  \citenamefont {Fowler}, \citenamefont {Raussendorf}, \citenamefont {Chen},
  \citenamefont {Liu}, \citenamefont {Lu}, \citenamefont {Deng},\ and\
  \citenamefont {Chen}}]{Yao12}%
  \BibitemOpen
  \bibfield  {author} {\bibinfo {author} {\bibfnamefont {X.-C.}\ \bibnamefont
  {Yao}}, \bibinfo {author} {\bibfnamefont {T.-X.}\ \bibnamefont {Wang}},
  \bibinfo {author} {\bibfnamefont {H.-Z.}\ \bibnamefont {Chen}}, \bibinfo
  {author} {\bibfnamefont {W.-B.}\ \bibnamefont {Gao}}, \bibinfo {author}
  {\bibfnamefont {A.~G.}\ \bibnamefont {Fowler}}, \bibinfo {author}
  {\bibfnamefont {R.}~\bibnamefont {Raussendorf}}, \bibinfo {author}
  {\bibfnamefont {Z.-B.}\ \bibnamefont {Chen}}, \bibinfo {author}
  {\bibfnamefont {N.-L.}\ \bibnamefont {Liu}}, \bibinfo {author} {\bibfnamefont
  {C.-Y.}\ \bibnamefont {Lu}}, \bibinfo {author} {\bibfnamefont {Y.-J.}\
  \bibnamefont {Deng}},\ and\ \bibinfo {author} {\bibfnamefont {J.-W.}\
  \bibnamefont {Chen}, \bibfnamefont {Yu-Aoand~Pan}},\ }\bibfield  {title}
  {\bibinfo {title} {Experimental demonstration of topological error
  correction},\ }\href {https://doi.org/10.1038/nature10770} {\bibfield
  {journal} {\bibinfo  {journal} {Nature}\ }\textbf {\bibinfo {volume} {482}},\
  \bibinfo {pages} {489} (\bibinfo {year} {2012}{\natexlab{a}})}\BibitemShut
  {NoStop}%
\bibitem [{\citenamefont {Terhal}(2015)}]{Terhal15}%
  \BibitemOpen
  \bibfield  {author} {\bibinfo {author} {\bibfnamefont {B.~M.}\ \bibnamefont
  {Terhal}},\ }\bibfield  {title} {\bibinfo {title} {Quantum error correction
  for quantum memories},\ }\href {https://doi.org/10.1103/RevModPhys.87.307}
  {\bibfield  {journal} {\bibinfo  {journal} {Rev. Mod. Phys.}\ }\textbf
  {\bibinfo {volume} {87}},\ \bibinfo {pages} {307} (\bibinfo {year}
  {2015})}\BibitemShut {NoStop}%
\bibitem [{\citenamefont {Temme}\ \emph {et~al.}(2017)\citenamefont {Temme},
  \citenamefont {Bravyi},\ and\ \citenamefont {Gambetta}}]{Temme17}%
  \BibitemOpen
  \bibfield  {author} {\bibinfo {author} {\bibfnamefont {K.}~\bibnamefont
  {Temme}}, \bibinfo {author} {\bibfnamefont {S.}~\bibnamefont {Bravyi}},\ and\
  \bibinfo {author} {\bibfnamefont {J.~M.}\ \bibnamefont {Gambetta}},\
  }\bibfield  {title} {\bibinfo {title} {Error mitigation for short-depth
  quantum circuits},\ }\href {https://doi.org/10.1103/PhysRevLett.119.180509}
  {\bibfield  {journal} {\bibinfo  {journal} {Phys. Rev. Lett.}\ }\textbf
  {\bibinfo {volume} {119}},\ \bibinfo {pages} {180509} (\bibinfo {year}
  {2017})}\BibitemShut {NoStop}%
\bibitem [{\citenamefont {Endo}\ \emph {et~al.}(2018)\citenamefont {Endo},
  \citenamefont {Benjamin},\ and\ \citenamefont {Li}}]{Endo18}%
  \BibitemOpen
  \bibfield  {author} {\bibinfo {author} {\bibfnamefont {S.}~\bibnamefont
  {Endo}}, \bibinfo {author} {\bibfnamefont {S.~C.}\ \bibnamefont {Benjamin}},\
  and\ \bibinfo {author} {\bibfnamefont {Y.}~\bibnamefont {Li}},\ }\bibfield
  {title} {\bibinfo {title} {Practical quantum error mitigation for near-future
  applications},\ }\href {https://doi.org/10.1103/PhysRevX.8.031027} {\bibfield
   {journal} {\bibinfo  {journal} {Phys. Rev. X}\ }\textbf {\bibinfo {volume}
  {8}},\ \bibinfo {pages} {031027} (\bibinfo {year} {2018})}\BibitemShut
  {NoStop}%
\bibitem [{\citenamefont {McArdle}\ \emph {et~al.}(2019)\citenamefont
  {McArdle}, \citenamefont {Yuan},\ and\ \citenamefont {Benjamin}}]{McArdle19}%
  \BibitemOpen
  \bibfield  {author} {\bibinfo {author} {\bibfnamefont {S.}~\bibnamefont
  {McArdle}}, \bibinfo {author} {\bibfnamefont {X.}~\bibnamefont {Yuan}},\ and\
  \bibinfo {author} {\bibfnamefont {S.}~\bibnamefont {Benjamin}},\ }\bibfield
  {title} {\bibinfo {title} {Error-mitigated digital quantum simulation},\
  }\href {https://doi.org/10.1103/PhysRevLett.122.180501} {\bibfield  {journal}
  {\bibinfo  {journal} {Phys. Rev. Lett.}\ }\textbf {\bibinfo {volume} {122}},\
  \bibinfo {pages} {180501} (\bibinfo {year} {2019})}\BibitemShut {NoStop}%
\bibitem [{\citenamefont {Kandala}\ \emph {et~al.}(2019)\citenamefont
  {Kandala}, \citenamefont {Temme}, \citenamefont {C\'orcoles}, \citenamefont
  {Mezzacapo}, \citenamefont {Chow},\ and\ \citenamefont
  {Gambetta}}]{Kandala19}%
  \BibitemOpen
  \bibfield  {author} {\bibinfo {author} {\bibfnamefont {A.}~\bibnamefont
  {Kandala}}, \bibinfo {author} {\bibfnamefont {K.}~\bibnamefont {Temme}},
  \bibinfo {author} {\bibfnamefont {A.~D.}\ \bibnamefont {C\'orcoles}},
  \bibinfo {author} {\bibfnamefont {A.}~\bibnamefont {Mezzacapo}}, \bibinfo
  {author} {\bibfnamefont {J.~M.}\ \bibnamefont {Chow}},\ and\ \bibinfo
  {author} {\bibfnamefont {J.~M.}\ \bibnamefont {Gambetta}},\ }\bibfield
  {title} {\bibinfo {title} {Error mitigation extends the computational reach
  of a noisy quantum processor},\ }\href
  {https://doi.org/10.1038/s41586-019-1040-7} {\bibfield  {journal} {\bibinfo
  {journal} {Nature}\ }\textbf {\bibinfo {volume} {567}},\ \bibinfo {pages}
  {491} (\bibinfo {year} {2019})}\BibitemShut {NoStop}%
\bibitem [{\citenamefont {Kwiat}\ \emph
  {et~al.}(2000{\natexlab{a}})\citenamefont {Kwiat}, \citenamefont {Berglund},
  \citenamefont {Altepeter},\ and\ \citenamefont {White}}]{Kwiat00}%
  \BibitemOpen
  \bibfield  {author} {\bibinfo {author} {\bibfnamefont {P.~G.}\ \bibnamefont
  {Kwiat}}, \bibinfo {author} {\bibfnamefont {A.~J.}\ \bibnamefont {Berglund}},
  \bibinfo {author} {\bibfnamefont {J.~B.}\ \bibnamefont {Altepeter}},\ and\
  \bibinfo {author} {\bibfnamefont {A.~G.}\ \bibnamefont {White}},\ }\bibfield
  {title} {\bibinfo {title} {Experimental verification of decoherence-free
  subspaces},\ }\href {https://doi.org/10.1126/science.290.5491.498} {\bibfield
   {journal} {\bibinfo  {journal} {Science}\ }\textbf {\bibinfo {volume}
  {290}},\ \bibinfo {pages} {498} (\bibinfo {year}
  {2000}{\natexlab{a}})}\BibitemShut {NoStop}%
\bibitem [{\citenamefont {Blume-Kohout}\ \emph {et~al.}(2008)\citenamefont
  {Blume-Kohout}, \citenamefont {Ng}, \citenamefont {Poulin},\ and\
  \citenamefont {Viola}}]{Blume08}%
  \BibitemOpen
  \bibfield  {author} {\bibinfo {author} {\bibfnamefont {R.}~\bibnamefont
  {Blume-Kohout}}, \bibinfo {author} {\bibfnamefont {H.~K.}\ \bibnamefont
  {Ng}}, \bibinfo {author} {\bibfnamefont {D.}~\bibnamefont {Poulin}},\ and\
  \bibinfo {author} {\bibfnamefont {L.}~\bibnamefont {Viola}},\ }\bibfield
  {title} {\bibinfo {title} {Characterizing the structure of preserved
  information in quantum processes},\ }\href
  {https://doi.org/10.1103/PhysRevLett.100.030501} {\bibfield  {journal}
  {\bibinfo  {journal} {Phys. Rev. Lett.}\ }\textbf {\bibinfo {volume} {100}},\
  \bibinfo {pages} {030501} (\bibinfo {year} {2008})}\BibitemShut {NoStop}%
\bibitem [{\citenamefont {Zanardi}\ and\ \citenamefont
  {Rasetti}(1997)}]{Zanardi97}%
  \BibitemOpen
  \bibfield  {author} {\bibinfo {author} {\bibfnamefont {P.}~\bibnamefont
  {Zanardi}}\ and\ \bibinfo {author} {\bibfnamefont {M.}~\bibnamefont
  {Rasetti}},\ }\bibfield  {title} {\bibinfo {title} {Noiseless quantum
  codes},\ }\href {https://doi.org/10.1103/PhysRevLett.79.3306} {\bibfield
  {journal} {\bibinfo  {journal} {Phys. Rev. Lett.}\ }\textbf {\bibinfo
  {volume} {79}},\ \bibinfo {pages} {3306} (\bibinfo {year}
  {1997})}\BibitemShut {NoStop}%
\bibitem [{\citenamefont {Lidar}\ \emph
  {et~al.}(1998{\natexlab{a}})\citenamefont {Lidar}, \citenamefont {Chuang},\
  and\ \citenamefont {Whaley}}]{Lidar98}%
  \BibitemOpen
  \bibfield  {author} {\bibinfo {author} {\bibfnamefont {D.~A.}\ \bibnamefont
  {Lidar}}, \bibinfo {author} {\bibfnamefont {I.~L.}\ \bibnamefont {Chuang}},\
  and\ \bibinfo {author} {\bibfnamefont {K.~B.}\ \bibnamefont {Whaley}},\
  }\bibfield  {title} {\bibinfo {title} {Decoherence-free subspaces for quantum
  computation},\ }\href {https://doi.org/10.1103/PhysRevLett.81.2594}
  {\bibfield  {journal} {\bibinfo  {journal} {Phys. Rev. Lett.}\ }\textbf
  {\bibinfo {volume} {81}},\ \bibinfo {pages} {2594} (\bibinfo {year}
  {1998}{\natexlab{a}})}\BibitemShut {NoStop}%
\bibitem [{\citenamefont {Bacon}\ \emph {et~al.}(2000)\citenamefont {Bacon},
  \citenamefont {Kempe}, \citenamefont {Lidar},\ and\ \citenamefont
  {Whaley}}]{Bacon00}%
  \BibitemOpen
  \bibfield  {author} {\bibinfo {author} {\bibfnamefont {D.}~\bibnamefont
  {Bacon}}, \bibinfo {author} {\bibfnamefont {J.}~\bibnamefont {Kempe}},
  \bibinfo {author} {\bibfnamefont {D.~A.}\ \bibnamefont {Lidar}},\ and\
  \bibinfo {author} {\bibfnamefont {K.~B.}\ \bibnamefont {Whaley}},\ }\bibfield
   {title} {\bibinfo {title} {Universal fault-tolerant quantum computation on
  decoherence-free subspaces},\ }\href
  {https://doi.org/10.1103/PhysRevLett.85.1758} {\bibfield  {journal} {\bibinfo
   {journal} {Phys. Rev. Lett.}\ }\textbf {\bibinfo {volume} {85}},\ \bibinfo
  {pages} {1758} (\bibinfo {year} {2000})}\BibitemShut {NoStop}%
\bibitem [{\citenamefont {Wu}\ \emph {et~al.}(2021)\citenamefont {Wu},
  \citenamefont {Zhou}, \citenamefont {Ye}, \citenamefont {Liu},\ and\
  \citenamefont {Yang}}]{Wu_2021}%
  \BibitemOpen
  \bibfield  {author} {\bibinfo {author} {\bibfnamefont {Q.-C.}\ \bibnamefont
  {Wu}}, \bibinfo {author} {\bibfnamefont {Y.-H.}\ \bibnamefont {Zhou}},
  \bibinfo {author} {\bibfnamefont {B.-L.}\ \bibnamefont {Ye}}, \bibinfo
  {author} {\bibfnamefont {T.}~\bibnamefont {Liu}},\ and\ \bibinfo {author}
  {\bibfnamefont {C.-P.}\ \bibnamefont {Yang}},\ }\bibfield  {title} {\bibinfo
  {title} {Nonadiabatic quantum state engineering by time-dependent
  decoherence-free subspaces in open quantum systems},\ }\href
  {https://doi.org/10.1088/1367-2630/ac309d} {\bibfield  {journal} {\bibinfo
  {journal} {New J. Phys.}\ }\textbf {\bibinfo {volume} {23}},\ \bibinfo
  {pages} {113005} (\bibinfo {year} {2021})}\BibitemShut {NoStop}%
\bibitem [{\citenamefont {Addis}\ \emph {et~al.}(2015)\citenamefont {Addis},
  \citenamefont {Ciccarello}, \citenamefont {Cascio}, \citenamefont {Palma},\
  and\ \citenamefont {Maniscalco}}]{Addis_2015}%
  \BibitemOpen
  \bibfield  {author} {\bibinfo {author} {\bibfnamefont {C.}~\bibnamefont
  {Addis}}, \bibinfo {author} {\bibfnamefont {F.}~\bibnamefont {Ciccarello}},
  \bibinfo {author} {\bibfnamefont {M.}~\bibnamefont {Cascio}}, \bibinfo
  {author} {\bibfnamefont {G.~M.}\ \bibnamefont {Palma}},\ and\ \bibinfo
  {author} {\bibfnamefont {S.}~\bibnamefont {Maniscalco}},\ }\bibfield  {title}
  {\bibinfo {title} {Dynamical decoupling efficiency versus quantum
  non-markovianity},\ }\href {https://doi.org/10.1088/1367-2630/17/12/123004}
  {\bibfield  {journal} {\bibinfo  {journal} {New J. Phys.}\ }\textbf {\bibinfo
  {volume} {17}},\ \bibinfo {pages} {123004} (\bibinfo {year}
  {2015})}\BibitemShut {NoStop}%
\bibitem [{\citenamefont {Burgarth}\ \emph {et~al.}(2021)\citenamefont
  {Burgarth}, \citenamefont {Facchi}, \citenamefont {Fraas},\ and\
  \citenamefont {Hillier}}]{Robin_Hillier2021}%
  \BibitemOpen
  \bibfield  {author} {\bibinfo {author} {\bibfnamefont {D.}~\bibnamefont
  {Burgarth}}, \bibinfo {author} {\bibfnamefont {P.}~\bibnamefont {Facchi}},
  \bibinfo {author} {\bibfnamefont {M.}~\bibnamefont {Fraas}},\ and\ \bibinfo
  {author} {\bibfnamefont {R.}~\bibnamefont {Hillier}},\ }\bibfield  {title}
  {\bibinfo {title} {{Non-Markovian noise that cannot be dynamically decoupled
  by periodic spin echo pulses}},\ }\href
  {https://doi.org/10.21468/SciPostPhys.11.2.027} {\bibfield  {journal}
  {\bibinfo  {journal} {SciPost Phys.}\ }\textbf {\bibinfo {volume} {11}},\
  \bibinfo {pages} {027} (\bibinfo {year} {2021})}\BibitemShut {NoStop}%
\bibitem [{\citenamefont {Lidar}\ \emph
  {et~al.}(1998{\natexlab{b}})\citenamefont {Lidar}, \citenamefont {Chuang},\
  and\ \citenamefont {Whaley}}]{Lidar1998}%
  \BibitemOpen
  \bibfield  {author} {\bibinfo {author} {\bibfnamefont {D.~A.}\ \bibnamefont
  {Lidar}}, \bibinfo {author} {\bibfnamefont {I.~L.}\ \bibnamefont {Chuang}},\
  and\ \bibinfo {author} {\bibfnamefont {K.~B.}\ \bibnamefont {Whaley}},\
  }\bibfield  {title} {\bibinfo {title} {Decoherence-free subspaces for quantum
  computation},\ }\href {https://doi.org/10.1103/PhysRevLett.81.2594}
  {\bibfield  {journal} {\bibinfo  {journal} {Phys. Rev. Lett.}\ }\textbf
  {\bibinfo {volume} {81}},\ \bibinfo {pages} {2594} (\bibinfo {year}
  {1998}{\natexlab{b}})}\BibitemShut {NoStop}%
\bibitem [{\citenamefont {Kwiat}\ \emph
  {et~al.}(2000{\natexlab{b}})\citenamefont {Kwiat}, \citenamefont {Berglund},
  \citenamefont {Altepeter},\ and\ \citenamefont {White}}]{Paul_2000}%
  \BibitemOpen
  \bibfield  {author} {\bibinfo {author} {\bibfnamefont {P.~G.}\ \bibnamefont
  {Kwiat}}, \bibinfo {author} {\bibfnamefont {A.~J.}\ \bibnamefont {Berglund}},
  \bibinfo {author} {\bibfnamefont {J.~B.}\ \bibnamefont {Altepeter}},\ and\
  \bibinfo {author} {\bibfnamefont {A.~G.}\ \bibnamefont {White}},\ }\bibfield
  {title} {\bibinfo {title} {Experimental verification of decoherence-free
  subspaces},\ }\href {https://doi.org/10.1126/science.290.5491.498} {\bibfield
   {journal} {\bibinfo  {journal} {Science}\ }\textbf {\bibinfo {volume}
  {290}},\ \bibinfo {pages} {498} (\bibinfo {year}
  {2000}{\natexlab{b}})}\BibitemShut {NoStop}%
\bibitem [{\citenamefont {He}\ \emph {et~al.}(2013)\citenamefont {He},
  \citenamefont {Yao},\ and\ \citenamefont {Zou}}]{He13}%
  \BibitemOpen
  \bibfield  {author} {\bibinfo {author} {\bibfnamefont {Z.}~\bibnamefont
  {He}}, \bibinfo {author} {\bibfnamefont {C.}~\bibnamefont {Yao}},\ and\
  \bibinfo {author} {\bibfnamefont {J.}~\bibnamefont {Zou}},\ }\bibfield
  {title} {\bibinfo {title} {Robust state transfer in the quantum spin channel
  via weak measurement and quantum measurement reversal},\ }\href
  {https://doi.org/10.1103/PhysRevA.88.044304} {\bibfield  {journal} {\bibinfo
  {journal} {Phys. Rev. A}\ }\textbf {\bibinfo {volume} {88}},\ \bibinfo
  {pages} {044304} (\bibinfo {year} {2013})}\BibitemShut {NoStop}%
\bibitem [{\citenamefont {Doustimotlagh}\ \emph {et~al.}(2014)\citenamefont
  {Doustimotlagh}, \citenamefont {Wang}, \citenamefont {You},\ and\
  \citenamefont {Long}}]{Doustimotlagh_2014}%
  \BibitemOpen
  \bibfield  {author} {\bibinfo {author} {\bibfnamefont {N.}~\bibnamefont
  {Doustimotlagh}}, \bibinfo {author} {\bibfnamefont {S.}~\bibnamefont {Wang}},
  \bibinfo {author} {\bibfnamefont {C.}~\bibnamefont {You}},\ and\ \bibinfo
  {author} {\bibfnamefont {G.-L.}\ \bibnamefont {Long}},\ }\bibfield  {title}
  {\bibinfo {title} {Enhancement of quantum correlations between two particles
  under decoherence in finite-temperature environment},\ }\href
  {https://doi.org/10.1209/0295-5075/106/60003} {\bibfield  {journal} {\bibinfo
   {journal} {{EPL} (Europhysics Letters)}\ }\textbf {\bibinfo {volume}
  {106}},\ \bibinfo {pages} {60003} (\bibinfo {year} {2014})}\BibitemShut
  {NoStop}%
\bibitem [{\citenamefont {Kim}\ \emph {et~al.}(2009)\citenamefont {Kim},
  \citenamefont {Cho}, \citenamefont {Ra},\ and\ \citenamefont {Kim}}]{Kim:09}%
  \BibitemOpen
  \bibfield  {author} {\bibinfo {author} {\bibfnamefont {Y.-S.}\ \bibnamefont
  {Kim}}, \bibinfo {author} {\bibfnamefont {Y.-W.}\ \bibnamefont {Cho}},
  \bibinfo {author} {\bibfnamefont {Y.-S.}\ \bibnamefont {Ra}},\ and\ \bibinfo
  {author} {\bibfnamefont {Y.-H.}\ \bibnamefont {Kim}},\ }\bibfield  {title}
  {\bibinfo {title} {Reversing the weak quantum measurement for a photonic
  qubit},\ }\href {https://doi.org/10.1364/OE.17.011978} {\bibfield  {journal}
  {\bibinfo  {journal} {Opt. Express}\ }\textbf {\bibinfo {volume} {17}},\
  \bibinfo {pages} {11978} (\bibinfo {year} {2009})}\BibitemShut {NoStop}%
\bibitem [{\citenamefont {Lim}\ \emph {et~al.}(2014)\citenamefont {Lim},
  \citenamefont {Lee}, \citenamefont {Hong},\ and\ \citenamefont
  {Kim}}]{Lim:14}%
  \BibitemOpen
  \bibfield  {author} {\bibinfo {author} {\bibfnamefont {H.-T.}\ \bibnamefont
  {Lim}}, \bibinfo {author} {\bibfnamefont {J.-C.}\ \bibnamefont {Lee}},
  \bibinfo {author} {\bibfnamefont {K.-H.}\ \bibnamefont {Hong}},\ and\
  \bibinfo {author} {\bibfnamefont {Y.-H.}\ \bibnamefont {Kim}},\ }\bibfield
  {title} {\bibinfo {title} {Avoiding entanglement sudden death using
  single-qubit quantum measurement reversal},\ }\href
  {https://doi.org/10.1364/OE.22.019055} {\bibfield  {journal} {\bibinfo
  {journal} {Opt. Express}\ }\textbf {\bibinfo {volume} {22}},\ \bibinfo
  {pages} {19055} (\bibinfo {year} {2014})}\BibitemShut {NoStop}%
\bibitem [{\citenamefont {Kim}\ \emph {et~al.}(2012)\citenamefont {Kim},
  \citenamefont {Lee}, \citenamefont {Kwon},\ and\ \citenamefont
  {Kim}}]{kim2012protecting}%
  \BibitemOpen
  \bibfield  {author} {\bibinfo {author} {\bibfnamefont {Y.-S.}\ \bibnamefont
  {Kim}}, \bibinfo {author} {\bibfnamefont {J.-C.}\ \bibnamefont {Lee}},
  \bibinfo {author} {\bibfnamefont {O.}~\bibnamefont {Kwon}},\ and\ \bibinfo
  {author} {\bibfnamefont {Y.-H.}\ \bibnamefont {Kim}},\ }\bibfield  {title}
  {\bibinfo {title} {Protecting entanglement from decoherence using weak
  measurement and quantum measurement reversal},\ }\href
  {https://doi.org/10.1038/nphys2178} {\bibfield  {journal} {\bibinfo
  {journal} {Nat. Phys.}\ }\textbf {\bibinfo {volume} {8}},\ \bibinfo {pages}
  {117} (\bibinfo {year} {2012})}\BibitemShut {NoStop}%
\bibitem [{\citenamefont {Sun}\ \emph {et~al.}(2010)\citenamefont {Sun},
  \citenamefont {Al-Amri}, \citenamefont {Davidovich},\ and\ \citenamefont
  {Zubairy}}]{Sun2010}%
  \BibitemOpen
  \bibfield  {author} {\bibinfo {author} {\bibfnamefont {Q.}~\bibnamefont
  {Sun}}, \bibinfo {author} {\bibfnamefont {M.}~\bibnamefont {Al-Amri}},
  \bibinfo {author} {\bibfnamefont {L.}~\bibnamefont {Davidovich}},\ and\
  \bibinfo {author} {\bibfnamefont {M.~S.}\ \bibnamefont {Zubairy}},\
  }\bibfield  {title} {\bibinfo {title} {Reversing entanglement change by a
  weak measurement},\ }\href {https://doi.org/10.1103/PhysRevA.82.052323}
  {\bibfield  {journal} {\bibinfo  {journal} {Phys. Rev. A}\ }\textbf {\bibinfo
  {volume} {82}},\ \bibinfo {pages} {052323} (\bibinfo {year}
  {2010})}\BibitemShut {NoStop}%
\bibitem [{\citenamefont {Zong}\ \emph {et~al.}(2014)\citenamefont {Zong},
  \citenamefont {Du}, \citenamefont {Yang}, \citenamefont {Yang},\ and\
  \citenamefont {Cao}}]{Zong2014}%
  \BibitemOpen
  \bibfield  {author} {\bibinfo {author} {\bibfnamefont {X.-L.}\ \bibnamefont
  {Zong}}, \bibinfo {author} {\bibfnamefont {C.-Q.}\ \bibnamefont {Du}},
  \bibinfo {author} {\bibfnamefont {M.}~\bibnamefont {Yang}}, \bibinfo {author}
  {\bibfnamefont {Q.}~\bibnamefont {Yang}},\ and\ \bibinfo {author}
  {\bibfnamefont {Z.-L.}\ \bibnamefont {Cao}},\ }\bibfield  {title} {\bibinfo
  {title} {Protecting remote bipartite entanglement against amplitude damping
  by local unitary operations},\ }\href
  {https://doi.org/10.1103/PhysRevA.90.062345} {\bibfield  {journal} {\bibinfo
  {journal} {Phys. Rev. A}\ }\textbf {\bibinfo {volume} {90}},\ \bibinfo
  {pages} {062345} (\bibinfo {year} {2014})}\BibitemShut {NoStop}%
\bibitem [{\citenamefont {Yao}\ \emph {et~al.}(2012{\natexlab{b}})\citenamefont
  {Yao}, \citenamefont {Ma}, \citenamefont {Chen},\ and\ \citenamefont
  {Serafini}}]{Yao2012}%
  \BibitemOpen
  \bibfield  {author} {\bibinfo {author} {\bibfnamefont {C.}~\bibnamefont
  {Yao}}, \bibinfo {author} {\bibfnamefont {Z.-H.}\ \bibnamefont {Ma}},
  \bibinfo {author} {\bibfnamefont {Z.-H.}\ \bibnamefont {Chen}},\ and\
  \bibinfo {author} {\bibfnamefont {A.}~\bibnamefont {Serafini}},\ }\bibfield
  {title} {\bibinfo {title} {Robust tripartite-to-bipartite entanglement
  localization by weak measurements and reversal},\ }\href
  {https://doi.org/10.1103/PhysRevA.86.022312} {\bibfield  {journal} {\bibinfo
  {journal} {Phys. Rev. A}\ }\textbf {\bibinfo {volume} {86}},\ \bibinfo
  {pages} {022312} (\bibinfo {year} {2012}{\natexlab{b}})}\BibitemShut
  {NoStop}%
\bibitem [{\citenamefont {Man}\ \emph {et~al.}(2012)\citenamefont {Man},
  \citenamefont {Xia},\ and\ \citenamefont {An}}]{Man2012}%
  \BibitemOpen
  \bibfield  {author} {\bibinfo {author} {\bibfnamefont {Z.-X.}\ \bibnamefont
  {Man}}, \bibinfo {author} {\bibfnamefont {Y.-J.}\ \bibnamefont {Xia}},\ and\
  \bibinfo {author} {\bibfnamefont {N.~B.}\ \bibnamefont {An}},\ }\bibfield
  {title} {\bibinfo {title} {Manipulating entanglement of two qubits in a
  common environment by means of weak measurements and quantum measurement
  reversals},\ }\href {https://doi.org/10.1103/PhysRevA.86.012325} {\bibfield
  {journal} {\bibinfo  {journal} {Phys. Rev. A}\ }\textbf {\bibinfo {volume}
  {86}},\ \bibinfo {pages} {012325} (\bibinfo {year} {2012})}\BibitemShut
  {NoStop}%
\bibitem [{\citenamefont {Starling}\ and\ \citenamefont
  {Williams}(2013)}]{Starling2013}%
  \BibitemOpen
  \bibfield  {author} {\bibinfo {author} {\bibfnamefont {D.~J.}\ \bibnamefont
  {Starling}}\ and\ \bibinfo {author} {\bibfnamefont {N.~S.}\ \bibnamefont
  {Williams}},\ }\bibfield  {title} {\bibinfo {title} {Efficacy of measurement
  reversal for stochastic disturbances},\ }\href
  {https://doi.org/10.1103/PhysRevA.88.024304} {\bibfield  {journal} {\bibinfo
  {journal} {Phys. Rev. A}\ }\textbf {\bibinfo {volume} {88}},\ \bibinfo
  {pages} {024304} (\bibinfo {year} {2013})}\BibitemShut {NoStop}%
\bibitem [{\citenamefont {Royer}(1994)}]{Royer1994}%
  \BibitemOpen
  \bibfield  {author} {\bibinfo {author} {\bibfnamefont {A.}~\bibnamefont
  {Royer}},\ }\bibfield  {title} {\bibinfo {title} {Reversible quantum
  measurements on a spin 1/2 and measuring the state of a single system},\
  }\href {https://doi.org/10.1103/PhysRevLett.73.913} {\bibfield  {journal}
  {\bibinfo  {journal} {Phys. Rev. Lett.}\ }\textbf {\bibinfo {volume} {73}},\
  \bibinfo {pages} {913} (\bibinfo {year} {1994})}\BibitemShut {NoStop}%
\bibitem [{\citenamefont {Korotkov}\ and\ \citenamefont
  {Jordan}(2006)}]{Korotkov2006}%
  \BibitemOpen
  \bibfield  {author} {\bibinfo {author} {\bibfnamefont {A.~N.}\ \bibnamefont
  {Korotkov}}\ and\ \bibinfo {author} {\bibfnamefont {A.~N.}\ \bibnamefont
  {Jordan}},\ }\bibfield  {title} {\bibinfo {title} {Undoing a weak quantum
  measurement of a solid-state qubit},\ }\href
  {https://doi.org/10.1103/PhysRevLett.97.166805} {\bibfield  {journal}
  {\bibinfo  {journal} {Phys. Rev. Lett.}\ }\textbf {\bibinfo {volume} {97}},\
  \bibinfo {pages} {166805} (\bibinfo {year} {2006})}\BibitemShut {NoStop}%
\bibitem [{\citenamefont {Sun}\ \emph {et~al.}(2009)\citenamefont {Sun},
  \citenamefont {Al-Amri},\ and\ \citenamefont {Zubairy}}]{Sun2009}%
  \BibitemOpen
  \bibfield  {author} {\bibinfo {author} {\bibfnamefont {Q.}~\bibnamefont
  {Sun}}, \bibinfo {author} {\bibfnamefont {M.}~\bibnamefont {Al-Amri}},\ and\
  \bibinfo {author} {\bibfnamefont {M.~S.}\ \bibnamefont {Zubairy}},\
  }\bibfield  {title} {\bibinfo {title} {Reversing the weak measurement of an
  arbitrary field with finite photon number},\ }\href
  {https://doi.org/10.1103/PhysRevA.80.033838} {\bibfield  {journal} {\bibinfo
  {journal} {Phys. Rev. A}\ }\textbf {\bibinfo {volume} {80}},\ \bibinfo
  {pages} {033838} (\bibinfo {year} {2009})}\BibitemShut {NoStop}%
\bibitem [{\citenamefont {Lee}\ \emph {et~al.}(2011)\citenamefont {Lee},
  \citenamefont {Jeong}, \citenamefont {Kim},\ and\ \citenamefont
  {Kim}}]{Lee:11}%
  \BibitemOpen
  \bibfield  {author} {\bibinfo {author} {\bibfnamefont {J.-C.}\ \bibnamefont
  {Lee}}, \bibinfo {author} {\bibfnamefont {Y.-C.}\ \bibnamefont {Jeong}},
  \bibinfo {author} {\bibfnamefont {Y.-S.}\ \bibnamefont {Kim}},\ and\ \bibinfo
  {author} {\bibfnamefont {Y.-H.}\ \bibnamefont {Kim}},\ }\bibfield  {title}
  {\bibinfo {title} {Experimental demonstration of decoherence suppression via
  quantum measurement reversal},\ }\href {https://doi.org/10.1364/OE.19.016309}
  {\bibfield  {journal} {\bibinfo  {journal} {Opt. Express}\ }\textbf {\bibinfo
  {volume} {19}},\ \bibinfo {pages} {16309} (\bibinfo {year}
  {2011})}\BibitemShut {NoStop}%
\bibitem [{\citenamefont {Singh}\ \emph {et~al.}(2023)\citenamefont {Singh},
  \citenamefont {Mazaheri}, \citenamefont {Peng}, \citenamefont {Sohail},
  \citenamefont {Khalid},\ and\ \citenamefont {Asjad}}]{singh2023}%
  \BibitemOpen
  \bibfield  {author} {\bibinfo {author} {\bibfnamefont {S.}~\bibnamefont
  {Singh}}, \bibinfo {author} {\bibfnamefont {M.}~\bibnamefont {Mazaheri}},
  \bibinfo {author} {\bibfnamefont {J.-X.}\ \bibnamefont {Peng}}, \bibinfo
  {author} {\bibfnamefont {A.}~\bibnamefont {Sohail}}, \bibinfo {author}
  {\bibfnamefont {M.}~\bibnamefont {Khalid}},\ and\ \bibinfo {author}
  {\bibfnamefont {M.}~\bibnamefont {Asjad}},\ }\bibfield  {title} {\bibinfo
  {title} {Enhanced weak force sensing based on atom-based coherent quantum
  noise cancellation in a hybrid cavity optomechanical system},\ }\href
  {https://doi.org/https://doi.org/10.3389/fphy.2023.1142452} {\bibfield
  {journal} {\bibinfo  {journal} {Front. Phys.}\ }\textbf {\bibinfo {volume}
  {11}},\ \bibinfo {pages} {245} (\bibinfo {year} {2023})}\BibitemShut
  {NoStop}%
\bibitem [{\citenamefont {Gelman}\ and\ \citenamefont
  {Mironov}(2010)}]{gelman2010}%
  \BibitemOpen
  \bibfield  {author} {\bibinfo {author} {\bibfnamefont {A.}~\bibnamefont
  {Gelman}}\ and\ \bibinfo {author} {\bibfnamefont {V.}~\bibnamefont
  {Mironov}},\ }\bibfield  {title} {\bibinfo {title} {Noise suppression in an
  atomic system under the action of a field in a squeezed coherent state},\
  }\href {https://doi.org/10.1134/S1063776110040011} {\bibfield  {journal}
  {\bibinfo  {journal} {J. Exp. Theor. Phys.}\ }\textbf {\bibinfo {volume}
  {110}},\ \bibinfo {pages} {551} (\bibinfo {year} {2010})}\BibitemShut
  {NoStop}%
\bibitem [{\citenamefont {Dantan}\ and\ \citenamefont
  {Pinard}(2004)}]{Dantan2004}%
  \BibitemOpen
  \bibfield  {author} {\bibinfo {author} {\bibfnamefont {A.}~\bibnamefont
  {Dantan}}\ and\ \bibinfo {author} {\bibfnamefont {M.}~\bibnamefont
  {Pinard}},\ }\bibfield  {title} {\bibinfo {title} {Quantum-state transfer
  between fields and atoms in electromagnetically induced transparency},\
  }\href {https://doi.org/10.1103/PhysRevA.69.043810} {\bibfield  {journal}
  {\bibinfo  {journal} {Phys. Rev. A}\ }\textbf {\bibinfo {volume} {69}},\
  \bibinfo {pages} {043810} (\bibinfo {year} {2004})}\BibitemShut {NoStop}%
\bibitem [{\citenamefont {Dantan}\ \emph {et~al.}(2005)\citenamefont {Dantan},
  \citenamefont {Bramati},\ and\ \citenamefont {Pinard}}]{Dantan2005}%
  \BibitemOpen
  \bibfield  {author} {\bibinfo {author} {\bibfnamefont {A.}~\bibnamefont
  {Dantan}}, \bibinfo {author} {\bibfnamefont {A.}~\bibnamefont {Bramati}},\
  and\ \bibinfo {author} {\bibfnamefont {M.}~\bibnamefont {Pinard}},\
  }\bibfield  {title} {\bibinfo {title} {Atomic quantum memory: Cavity versus
  single-pass schemes},\ }\href {https://doi.org/10.1103/PhysRevA.71.043801}
  {\bibfield  {journal} {\bibinfo  {journal} {Phys. Rev. A}\ }\textbf {\bibinfo
  {volume} {71}},\ \bibinfo {pages} {043801} (\bibinfo {year}
  {2005})}\BibitemShut {NoStop}%
\bibitem [{\citenamefont {Barberis-Blostein}\ and\ \citenamefont
  {Bienert}(2007)}]{Barberis2007}%
  \BibitemOpen
  \bibfield  {author} {\bibinfo {author} {\bibfnamefont {P.}~\bibnamefont
  {Barberis-Blostein}}\ and\ \bibinfo {author} {\bibfnamefont {M.}~\bibnamefont
  {Bienert}},\ }\bibfield  {title} {\bibinfo {title} {Opacity of
  electromagnetically induced transparency for quantum fluctuations},\ }\href
  {https://doi.org/10.1103/PhysRevLett.98.033602} {\bibfield  {journal}
  {\bibinfo  {journal} {Phys. Rev. Lett.}\ }\textbf {\bibinfo {volume} {98}},\
  \bibinfo {pages} {033602} (\bibinfo {year} {2007})}\BibitemShut {NoStop}%
\bibitem [{\citenamefont {Marangos}(1998)}]{J.P_1998}%
  \BibitemOpen
  \bibfield  {author} {\bibinfo {author} {\bibfnamefont {J.~P.}\ \bibnamefont
  {Marangos}},\ }\bibfield  {title} {\bibinfo {title} {Electromagnetically
  induced transparency},\ }\href {https://doi.org/10.1080/09500349808231909}
  {\bibfield  {journal} {\bibinfo  {journal} {J. Mod. Opt.}\ }\textbf {\bibinfo
  {volume} {45}},\ \bibinfo {pages} {471} (\bibinfo {year} {1998})}\BibitemShut
  {NoStop}%
\bibitem [{\citenamefont {Reiserer}\ \emph
  {et~al.}(2014{\natexlab{a}})\citenamefont {Reiserer}, \citenamefont {Kalb},
  \citenamefont {Rempe},\ and\ \citenamefont {Ritter}}]{reiserer2014quantum}%
  \BibitemOpen
  \bibfield  {author} {\bibinfo {author} {\bibfnamefont {A.}~\bibnamefont
  {Reiserer}}, \bibinfo {author} {\bibfnamefont {N.}~\bibnamefont {Kalb}},
  \bibinfo {author} {\bibfnamefont {G.}~\bibnamefont {Rempe}},\ and\ \bibinfo
  {author} {\bibfnamefont {S.}~\bibnamefont {Ritter}},\ }\bibfield  {title}
  {\bibinfo {title} {A quantum gate between a flying optical photon and a
  single trapped atom},\ }\href
  {https://doi.org/https://doi.org/10.1038/nature13177} {\bibfield  {journal}
  {\bibinfo  {journal} {Nature}\ }\textbf {\bibinfo {volume} {508}},\ \bibinfo
  {pages} {237} (\bibinfo {year} {2014}{\natexlab{a}})}\BibitemShut {NoStop}%
\bibitem [{\citenamefont {Hacker}\ \emph
  {et~al.}(2016{\natexlab{a}})\citenamefont {Hacker}, \citenamefont {Welte},
  \citenamefont {Rempe},\ and\ \citenamefont {Ritter}}]{hacker2016}%
  \BibitemOpen
  \bibfield  {author} {\bibinfo {author} {\bibfnamefont {B.}~\bibnamefont
  {Hacker}}, \bibinfo {author} {\bibfnamefont {S.}~\bibnamefont {Welte}},
  \bibinfo {author} {\bibfnamefont {G.}~\bibnamefont {Rempe}},\ and\ \bibinfo
  {author} {\bibfnamefont {S.}~\bibnamefont {Ritter}},\ }\bibfield  {title}
  {\bibinfo {title} {A photon--photon quantum gate based on a single atom in an
  optical resonator},\ }\href
  {https://doi.org/https://doi.org/10.1038/nature18592} {\bibfield  {journal}
  {\bibinfo  {journal} {Nature}\ }\textbf {\bibinfo {volume} {536}},\ \bibinfo
  {pages} {193} (\bibinfo {year} {2016}{\natexlab{a}})}\BibitemShut {NoStop}%
\bibitem [{\citenamefont {Kim}\ \emph {et~al.}(2013)\citenamefont {Kim},
  \citenamefont {Bose}, \citenamefont {Shen}, \citenamefont {Solomon},\ and\
  \citenamefont {Waks}}]{kim2013quantum}%
  \BibitemOpen
  \bibfield  {author} {\bibinfo {author} {\bibfnamefont {H.}~\bibnamefont
  {Kim}}, \bibinfo {author} {\bibfnamefont {R.}~\bibnamefont {Bose}}, \bibinfo
  {author} {\bibfnamefont {T.~C.}\ \bibnamefont {Shen}}, \bibinfo {author}
  {\bibfnamefont {G.~S.}\ \bibnamefont {Solomon}},\ and\ \bibinfo {author}
  {\bibfnamefont {E.}~\bibnamefont {Waks}},\ }\bibfield  {title} {\bibinfo
  {title} {A quantum logic gate between a solid-state quantum bit and a
  photon},\ }\href {https://doi.org/https://doi.org/10.1038/nphoton.2013.48}
  {\bibfield  {journal} {\bibinfo  {journal} {Nat. Photon.}\ }\textbf {\bibinfo
  {volume} {7}},\ \bibinfo {pages} {373} (\bibinfo {year} {2013})}\BibitemShut
  {NoStop}%
\bibitem [{\citenamefont {Duan}\ and\ \citenamefont
  {Kimble}(2004)}]{duan2004scalable}%
  \BibitemOpen
  \bibfield  {author} {\bibinfo {author} {\bibfnamefont {L.-M.}\ \bibnamefont
  {Duan}}\ and\ \bibinfo {author} {\bibfnamefont {H.~J.}\ \bibnamefont
  {Kimble}},\ }\bibfield  {title} {\bibinfo {title} {Scalable photonic quantum
  computation through cavity-assisted interactions},\ }\href
  {https://doi.org/10.1103/PhysRevLett.92.127902} {\bibfield  {journal}
  {\bibinfo  {journal} {Phys. Rev. Lett.}\ }\textbf {\bibinfo {volume} {92}},\
  \bibinfo {pages} {127902} (\bibinfo {year} {2004})}\BibitemShut {NoStop}%
\bibitem [{\citenamefont {Wang}\ \emph {et~al.}(2019)\citenamefont {Wang},
  \citenamefont {Li}, \citenamefont {Zhang}, \citenamefont {Su}, \citenamefont
  {Zhou}, \citenamefont {Liao}, \citenamefont {Du}, \citenamefont {Yan},\ and\
  \citenamefont {Zhu}}]{wang19}%
  \BibitemOpen
  \bibfield  {author} {\bibinfo {author} {\bibfnamefont {Y.}~\bibnamefont
  {Wang}}, \bibinfo {author} {\bibfnamefont {J.}~\bibnamefont {Li}}, \bibinfo
  {author} {\bibfnamefont {S.}~\bibnamefont {Zhang}}, \bibinfo {author}
  {\bibfnamefont {K.}~\bibnamefont {Su}}, \bibinfo {author} {\bibfnamefont
  {Y.}~\bibnamefont {Zhou}}, \bibinfo {author} {\bibfnamefont {K.}~\bibnamefont
  {Liao}}, \bibinfo {author} {\bibfnamefont {S.}~\bibnamefont {Du}}, \bibinfo
  {author} {\bibfnamefont {H.}~\bibnamefont {Yan}},\ and\ \bibinfo {author}
  {\bibfnamefont {S.-L.}\ \bibnamefont {Zhu}},\ }\bibfield  {title} {\bibinfo
  {title} {Efficient quantum memory for single-photon polarization qubits},\
  }\href {https://doi.org/https://doi.org/10.1038/s41566-019-0368-8} {\bibfield
   {journal} {\bibinfo  {journal} {Nat. Photon.}\ }\textbf {\bibinfo {volume}
  {13}},\ \bibinfo {pages} {346} (\bibinfo {year} {2019})}\BibitemShut
  {NoStop}%
\bibitem [{\citenamefont {Guo}\ \emph {et~al.}(2019)\citenamefont {Guo},
  \citenamefont {Feng}, \citenamefont {Yang}, \citenamefont {Yu}, \citenamefont
  {Chen}, \citenamefont {Yuan},\ and\ \citenamefont {Zhang}}]{guo19}%
  \BibitemOpen
  \bibfield  {author} {\bibinfo {author} {\bibfnamefont {J.}~\bibnamefont
  {Guo}}, \bibinfo {author} {\bibfnamefont {X.}~\bibnamefont {Feng}}, \bibinfo
  {author} {\bibfnamefont {P.}~\bibnamefont {Yang}}, \bibinfo {author}
  {\bibfnamefont {Z.}~\bibnamefont {Yu}}, \bibinfo {author} {\bibfnamefont
  {L.}~\bibnamefont {Chen}}, \bibinfo {author} {\bibfnamefont {C.-H.}\
  \bibnamefont {Yuan}},\ and\ \bibinfo {author} {\bibfnamefont
  {W.}~\bibnamefont {Zhang}},\ }\bibfield  {title} {\bibinfo {title}
  {High-performance raman quantum memory with optimal control in room
  temperature atoms},\ }\href
  {https://doi.org/https://doi.org/10.1038/s41467-018-08118-5} {\bibfield
  {journal} {\bibinfo  {journal} {Nat. Commun.}\ }\textbf {\bibinfo {volume}
  {10}},\ \bibinfo {pages} {148} (\bibinfo {year} {2019})}\BibitemShut
  {NoStop}%
\bibitem [{\citenamefont {Hosseini}\ \emph {et~al.}(2011)\citenamefont
  {Hosseini}, \citenamefont {Sparkes}, \citenamefont {Campbell}, \citenamefont
  {Lam},\ and\ \citenamefont {Buchler}}]{hosseini11}%
  \BibitemOpen
  \bibfield  {author} {\bibinfo {author} {\bibfnamefont {M.}~\bibnamefont
  {Hosseini}}, \bibinfo {author} {\bibfnamefont {B.~M.}\ \bibnamefont
  {Sparkes}}, \bibinfo {author} {\bibfnamefont {G.}~\bibnamefont {Campbell}},
  \bibinfo {author} {\bibfnamefont {P.~K.}\ \bibnamefont {Lam}},\ and\ \bibinfo
  {author} {\bibfnamefont {B.~C.}\ \bibnamefont {Buchler}},\ }\bibfield
  {title} {\bibinfo {title} {High efficiency coherent optical memory with warm
  rubidium vapour},\ }\href
  {https://doi.org/https://doi.org/10.1038/ncomms1175} {\bibfield  {journal}
  {\bibinfo  {journal} {Nat. Commun.}\ }\textbf {\bibinfo {volume} {2}},\
  \bibinfo {pages} {174} (\bibinfo {year} {2011})}\BibitemShut {NoStop}%
\bibitem [{\citenamefont {Luo}\ \emph {et~al.}(2015)\citenamefont {Luo},
  \citenamefont {Wang}, \citenamefont {Tong},\ and\ \citenamefont
  {Wang}}]{luo2015}%
  \BibitemOpen
  \bibfield  {author} {\bibinfo {author} {\bibfnamefont {S.}~\bibnamefont
  {Luo}}, \bibinfo {author} {\bibfnamefont {Y.}~\bibnamefont {Wang}}, \bibinfo
  {author} {\bibfnamefont {X.}~\bibnamefont {Tong}},\ and\ \bibinfo {author}
  {\bibfnamefont {Z.}~\bibnamefont {Wang}},\ }\bibfield  {title} {\bibinfo
  {title} {Graphene-based optical modulators},\ }\href
  {https://doi.org/https://doi.org/10.1186/s11671-015-0866-7} {\bibfield
  {journal} {\bibinfo  {journal} {Nanoscale Res. Lett.}\ }\textbf {\bibinfo
  {volume} {10}},\ \bibinfo {pages} {1} (\bibinfo {year} {2015})}\BibitemShut
  {NoStop}%
\bibitem [{\citenamefont {Amin}\ \emph {et~al.}(2018)\citenamefont {Amin},
  \citenamefont {Khurgin},\ and\ \citenamefont {Sorger}}]{amin2018}%
  \BibitemOpen
  \bibfield  {author} {\bibinfo {author} {\bibfnamefont {R.}~\bibnamefont
  {Amin}}, \bibinfo {author} {\bibfnamefont {J.~B.}\ \bibnamefont {Khurgin}},\
  and\ \bibinfo {author} {\bibfnamefont {V.~J.}\ \bibnamefont {Sorger}},\
  }\bibfield  {title} {\bibinfo {title} {Waveguide-based electro-absorption
  modulator performance: comparative analysis},\ }\href
  {https://doi.org/https://doi.org/10.1364/OE.26.015445} {\bibfield  {journal}
  {\bibinfo  {journal} {Opt. Express}\ }\textbf {\bibinfo {volume} {26}},\
  \bibinfo {pages} {15445} (\bibinfo {year} {2018})}\BibitemShut {NoStop}%
\bibitem [{\citenamefont {Baksic}\ \emph {et~al.}(2016)\citenamefont {Baksic},
  \citenamefont {Ribeiro},\ and\ \citenamefont {Clerk}}]{baksic2016}%
  \BibitemOpen
  \bibfield  {author} {\bibinfo {author} {\bibfnamefont {A.}~\bibnamefont
  {Baksic}}, \bibinfo {author} {\bibfnamefont {H.}~\bibnamefont {Ribeiro}},\
  and\ \bibinfo {author} {\bibfnamefont {A.~A.}\ \bibnamefont {Clerk}},\
  }\bibfield  {title} {\bibinfo {title} {Speeding up adiabatic quantum state
  transfer by using dressed states},\ }\href
  {https://doi.org/10.1103/PhysRevLett.116.230503} {\bibfield  {journal}
  {\bibinfo  {journal} {Phys. Rev. Lett.}\ }\textbf {\bibinfo {volume} {116}},\
  \bibinfo {pages} {230503} (\bibinfo {year} {2016})}\BibitemShut {NoStop}%
\bibitem [{\citenamefont {Chen}\ \emph {et~al.}(2010)\citenamefont {Chen},
  \citenamefont {Lizuain}, \citenamefont {Ruschhaupt}, \citenamefont
  {Gu\'ery-Odelin},\ and\ \citenamefont {Muga}}]{chen2010}%
  \BibitemOpen
  \bibfield  {author} {\bibinfo {author} {\bibfnamefont {X.}~\bibnamefont
  {Chen}}, \bibinfo {author} {\bibfnamefont {I.}~\bibnamefont {Lizuain}},
  \bibinfo {author} {\bibfnamefont {A.}~\bibnamefont {Ruschhaupt}}, \bibinfo
  {author} {\bibfnamefont {D.}~\bibnamefont {Gu\'ery-Odelin}},\ and\ \bibinfo
  {author} {\bibfnamefont {J.~G.}\ \bibnamefont {Muga}},\ }\bibfield  {title}
  {\bibinfo {title} {Shortcut to adiabatic passage in two- and three-level
  atoms},\ }\href {https://doi.org/10.1103/PhysRevLett.105.123003} {\bibfield
  {journal} {\bibinfo  {journal} {Phys. Rev. Lett.}\ }\textbf {\bibinfo
  {volume} {105}},\ \bibinfo {pages} {123003} (\bibinfo {year}
  {2010})}\BibitemShut {NoStop}%
\bibitem [{\citenamefont {Zhou}\ \emph {et~al.}(2017)\citenamefont {Zhou},
  \citenamefont {Baksic}, \citenamefont {Ribeiro}, \citenamefont {Yale},
  \citenamefont {Heremans}, \citenamefont {Jerger}, \citenamefont {Auer},
  \citenamefont {Burkard}, \citenamefont {Clerk},\ and\ \citenamefont
  {Awschalom}}]{zhou2017}%
  \BibitemOpen
  \bibfield  {author} {\bibinfo {author} {\bibfnamefont {B.~B.}\ \bibnamefont
  {Zhou}}, \bibinfo {author} {\bibfnamefont {A.}~\bibnamefont {Baksic}},
  \bibinfo {author} {\bibfnamefont {H.}~\bibnamefont {Ribeiro}}, \bibinfo
  {author} {\bibfnamefont {C.~G.}\ \bibnamefont {Yale}}, \bibinfo {author}
  {\bibfnamefont {F.~J.}\ \bibnamefont {Heremans}}, \bibinfo {author}
  {\bibfnamefont {P.~C.}\ \bibnamefont {Jerger}}, \bibinfo {author}
  {\bibfnamefont {A.}~\bibnamefont {Auer}}, \bibinfo {author} {\bibfnamefont
  {G.}~\bibnamefont {Burkard}}, \bibinfo {author} {\bibfnamefont {A.~A.}\
  \bibnamefont {Clerk}},\ and\ \bibinfo {author} {\bibfnamefont {D.~D.}\
  \bibnamefont {Awschalom}},\ }\bibfield  {title} {\bibinfo {title}
  {Accelerated quantum control using superadiabatic dynamics in a solid-state
  lambda system},\ }\href {https://doi.org/10.1038/nphys3967} {\bibfield
  {journal} {\bibinfo  {journal} {Nat. Phys.}\ }\textbf {\bibinfo {volume}
  {13}},\ \bibinfo {pages} {330} (\bibinfo {year} {2017})}\BibitemShut
  {NoStop}%
\bibitem [{\citenamefont {Vitanov}\ \emph {et~al.}(2017)\citenamefont
  {Vitanov}, \citenamefont {Rangelov}, \citenamefont {Shore},\ and\
  \citenamefont {Bergmann}}]{Vitanov_2017}%
  \BibitemOpen
  \bibfield  {author} {\bibinfo {author} {\bibfnamefont {N.~V.}\ \bibnamefont
  {Vitanov}}, \bibinfo {author} {\bibfnamefont {A.~A.}\ \bibnamefont
  {Rangelov}}, \bibinfo {author} {\bibfnamefont {B.~W.}\ \bibnamefont
  {Shore}},\ and\ \bibinfo {author} {\bibfnamefont {K.}~\bibnamefont
  {Bergmann}},\ }\bibfield  {title} {\bibinfo {title} {Stimulated raman
  adiabatic passage in physics, chemistry, and beyond},\ }\href
  {https://doi.org/10.1103/RevModPhys.89.015006} {\bibfield  {journal}
  {\bibinfo  {journal} {Rev. Mod. Phys.}\ }\textbf {\bibinfo {volume} {89}},\
  \bibinfo {pages} {015006} (\bibinfo {year} {2017})}\BibitemShut {NoStop}%
\bibitem [{\citenamefont {Wu}\ \emph {et~al.}(2022)\citenamefont {Wu},
  \citenamefont {Zhou}, \citenamefont {Ye}, \citenamefont {Liu}, \citenamefont
  {Zhao}, \citenamefont {Chen},\ and\ \citenamefont {Yang}}]{wu2022}%
  \BibitemOpen
  \bibfield  {author} {\bibinfo {author} {\bibfnamefont {Q.-C.}\ \bibnamefont
  {Wu}}, \bibinfo {author} {\bibfnamefont {Y.-H.}\ \bibnamefont {Zhou}},
  \bibinfo {author} {\bibfnamefont {B.-L.}\ \bibnamefont {Ye}}, \bibinfo
  {author} {\bibfnamefont {T.}~\bibnamefont {Liu}}, \bibinfo {author}
  {\bibfnamefont {J.-L.}\ \bibnamefont {Zhao}}, \bibinfo {author}
  {\bibfnamefont {D.-X.}\ \bibnamefont {Chen}},\ and\ \bibinfo {author}
  {\bibfnamefont {C.-P.}\ \bibnamefont {Yang}},\ }\bibfield  {title} {\bibinfo
  {title} {Generation of an enhanced multi-mode optomechanical-like quantum
  system and its application in creating hybrid entangled states},\ }\href
  {https://doi.org/https://doi.org/10.1002/andp.202100393} {\bibfield
  {journal} {\bibinfo  {journal} {Ann. Phys.}\ }\textbf {\bibinfo {volume}
  {534}},\ \bibinfo {pages} {2100393} (\bibinfo {year} {2022})}\BibitemShut
  {NoStop}%
\bibitem [{\citenamefont {Hacker}\ \emph {et~al.}(2019)\citenamefont {Hacker},
  \citenamefont {Welte}, \citenamefont {Daiss}, \citenamefont {Shaukat},
  \citenamefont {Ritter}, \citenamefont {Li},\ and\ \citenamefont
  {Rempe}}]{hacker2019}%
  \BibitemOpen
  \bibfield  {author} {\bibinfo {author} {\bibfnamefont {B.}~\bibnamefont
  {Hacker}}, \bibinfo {author} {\bibfnamefont {S.}~\bibnamefont {Welte}},
  \bibinfo {author} {\bibfnamefont {S.}~\bibnamefont {Daiss}}, \bibinfo
  {author} {\bibfnamefont {A.}~\bibnamefont {Shaukat}}, \bibinfo {author}
  {\bibfnamefont {S.}~\bibnamefont {Ritter}}, \bibinfo {author} {\bibfnamefont
  {L.}~\bibnamefont {Li}},\ and\ \bibinfo {author} {\bibfnamefont
  {G.}~\bibnamefont {Rempe}},\ }\bibfield  {title} {\bibinfo {title}
  {Deterministic creation of entangled atom-light schr{\"o}dinger-cat states},\
  }\href {https://doi.org/10.1038/s41566-018-0339-5} {\bibfield  {journal}
  {\bibinfo  {journal} {Nat. Photon.}\ }\textbf {\bibinfo {volume} {13}},\
  \bibinfo {pages} {110} (\bibinfo {year} {2019})}\BibitemShut {NoStop}%
\bibitem [{\citenamefont {Reiserer}\ \emph
  {et~al.}(2014{\natexlab{b}})\citenamefont {Reiserer}, \citenamefont {Kalb},
  \citenamefont {Rempe},\ and\ \citenamefont {Ritter}}]{Reiserer14}%
  \BibitemOpen
  \bibfield  {author} {\bibinfo {author} {\bibfnamefont {A.}~\bibnamefont
  {Reiserer}}, \bibinfo {author} {\bibfnamefont {N.}~\bibnamefont {Kalb}},
  \bibinfo {author} {\bibfnamefont {G.}~\bibnamefont {Rempe}},\ and\ \bibinfo
  {author} {\bibfnamefont {S.}~\bibnamefont {Ritter}},\ }\bibfield  {title}
  {\bibinfo {title} {Photon-mediated quantum gate between two neutral atoms in
  an optical cavity},\ }\href {https://doi.org/10.1038/nature13177} {\bibfield
  {journal} {\bibinfo  {journal} {Nature}\ }\textbf {\bibinfo {volume} {508}},\
  \bibinfo {pages} {237} (\bibinfo {year} {2014}{\natexlab{b}})}\BibitemShut
  {NoStop}%
\bibitem [{\citenamefont {Hacker}\ \emph
  {et~al.}(2016{\natexlab{b}})\citenamefont {Hacker}, \citenamefont {Welte},
  \citenamefont {Rempe},\ and\ \citenamefont {Ritter}}]{Hacker16}%
  \BibitemOpen
  \bibfield  {author} {\bibinfo {author} {\bibfnamefont {B.}~\bibnamefont
  {Hacker}}, \bibinfo {author} {\bibfnamefont {S.}~\bibnamefont {Welte}},
  \bibinfo {author} {\bibfnamefont {G.}~\bibnamefont {Rempe}},\ and\ \bibinfo
  {author} {\bibfnamefont {S.}~\bibnamefont {Ritter}},\ }\bibfield  {title}
  {\bibinfo {title} {A photon-photon quantum gate based on a single atom in an
  optical resonator},\ }\href {https://www.nature.com/articles/nature18592}
  {\bibfield  {journal} {\bibinfo  {journal} {Nature}\ }\textbf {\bibinfo
  {volume} {536}},\ \bibinfo {pages} {193} (\bibinfo {year}
  {2016}{\natexlab{b}})}\BibitemShut {NoStop}%
\bibitem [{\citenamefont {Welte}\ \emph {et~al.}(2018)\citenamefont {Welte},
  \citenamefont {Hacker}, \citenamefont {Daiss}, \citenamefont {Ritter},\ and\
  \citenamefont {Rempe}}]{Welte18}%
  \BibitemOpen
  \bibfield  {author} {\bibinfo {author} {\bibfnamefont {S.}~\bibnamefont
  {Welte}}, \bibinfo {author} {\bibfnamefont {B.}~\bibnamefont {Hacker}},
  \bibinfo {author} {\bibfnamefont {S.}~\bibnamefont {Daiss}}, \bibinfo
  {author} {\bibfnamefont {S.}~\bibnamefont {Ritter}},\ and\ \bibinfo {author}
  {\bibfnamefont {G.}~\bibnamefont {Rempe}},\ }\bibfield  {title} {\bibinfo
  {title} {Photon-mediated quantum gate between two neutral atoms in an optical
  cavity},\ }\href {https://doi.org/10.1103/PhysRevX.8.011018} {\bibfield
  {journal} {\bibinfo  {journal} {Phys. Rev. X}\ }\textbf {\bibinfo {volume}
  {8}},\ \bibinfo {pages} {011018} (\bibinfo {year} {2018})}\BibitemShut
  {NoStop}%
\bibitem [{\citenamefont {Benenti}\ and\ \citenamefont
  {Strini}(2010)}]{Benenti_2010}%
  \BibitemOpen
  \bibfield  {author} {\bibinfo {author} {\bibfnamefont {G.}~\bibnamefont
  {Benenti}}\ and\ \bibinfo {author} {\bibfnamefont {G.}~\bibnamefont
  {Strini}},\ }\bibfield  {title} {\bibinfo {title} {Computing the distance
  between quantum channels: usefulness of the fano representation},\ }\href
  {https://doi.org/10.1088/0953-4075/43/21/215508} {\bibfield  {journal}
  {\bibinfo  {journal} {J. Phys. B: At. Mol. Opt. Phys.}\ }\textbf {\bibinfo
  {volume} {43}},\ \bibinfo {pages} {215508} (\bibinfo {year}
  {2010})}\BibitemShut {NoStop}%
\end{thebibliography}
%

	\onecolumngrid
	
	\appendix

	\section{The protocol for qubit channels}
	\label{app:Qubit calculation}
	Detailed derivation of the protocol for arbitrary qubit channels. The protocol is implemented in five steps, to make arbitrary qubit channel $\Lambda^S$ transparent, i.e., $\Lambda^S \to \mathds{1}^S$, where $\mathds{1}^S$ is an identity channel.  The protocol uses two qubit ancilla $AB$ with is also allowed to undergo a wide class of noisy operation during the protocol, as discussed below. 
	
	\begin{enumerate}
		
		\item[Step 1:] To start with we attach the ancilla $AB$ with the system $S$ in an initial state
		\begin{align}
			\rho^{ABS}_1=\proj{++}^{AB}\otimes\rho^S,
		\end{align}
		where $\ket{\pm}^X=\frac{1}{\sqrt{2}}\Big(\ket{0}^X\pm\ket{1}^X\Big)$ with $X=A, B$, and $\rho^S$ is any state of $S$.

		\item[Step 2:] Initial state evolve with a global unitary (acausal operation) $U_{ABS}$.
		\begin{align}
			\rho_2^{ABS}=U_{ABS}\rho_1^{ABS}U_{ABS}^\dagger,
		\end{align}
		where $U_{ABS}=\proj{00}^{AB} \otimes  \mathds{1}^S + \proj{01}^{AB} \otimes \sigma_z^S+ \proj{10}^{AB} \otimes \sigma_x^S - i \proj{11}^{AB} \otimes \sigma^S_y$. Here $\{\sigma_x, \ \sigma_y, \ \sigma_z\}$ are the Pauli matrices.

		\item[Step 3:] System $S$ passes through a noisy channel implementing an arbitrary quantum noisy channel (CPTP map) $\Lambda^S$ that we want to eliminate, given by $\Lambda^{S}(\rho_{S})=\sum_{m}F_{m}^{S}\rho_{S}{F^{S}_{m}}^\dagger$, with Kraus operators $F_m^S=\sum_{i=0}^3 c_{mi} \sigma_i$ where $c_{mi}\in \mathds{C}$ can take arbitrary values satisfying $ \sum_m {F^S}^\dagger_mF_m^S=\mathds{1}^S$. Here we assume $\sigma_0=\mathds{1}$. The ancilla $AB$ may experience a class of environmental noise, given by the channels $\Phi^{AB}(\rho_{AB})=\sum_{\mu}E_{\mu}^{AB}\rho_{AB}{E^{AB}_{\mu}}^\dagger$, where the Karus operators are $E_\mu^{AB}=\sum_{j,k=0}^1q_{\mu jk} \sigma_z^j\sigma_x^k\otimes \sigma_x^{2-j}= q_{\mu 00} \mathds{1} \otimes \mathds{1} + 	q_{\mu 01} \sigma_x \otimes \mathds{1} + q_{\mu 10} \sigma_z \otimes \sigma_x + q_{\mu 11} \sigma_z\sigma_x \otimes \sigma_x$, satisfying $\sum_\mu {E^{AB}_\mu}^\dagger E_\mu^{AB}=\mathds{1}^{AB}$.  As a consequence, the overall state evolves as
		\begin{align}
			\rho_3^{ABS}=\Phi^{AB}\otimes\Lambda^S\Big(\rho_2^{ABS}\Big),
		\end{align} 
		The we choose $E_\mu^{AB}$, because of decoherence free subspace as discussed in the main text.
		\item[Step 4:] Now, $\rho_3^{ABS}$ is evolved with hermitian conjugate on the global unitary (acausal operation) $U_{ABS}^\dagger$. Thus,
		\begin{align}
			\rho_4^{ABS}=U_{ABS}^\dagger\rho_3^{ABS}U_{ABS},
		\end{align}
		where $U_{ABS}^\dagger=\proj{00}^{AB} \otimes  \mathds{1}^S + \proj{01}^{AB} \otimes \sigma_z^S + \proj{10}^{AB} \otimes \sigma_x^S + i \proj{11}^{AB} \otimes \sigma^S_y$. The state $\rho_4^{ABS}$ can explicitly be expressed as, 
		\begin{align}
			\rho_4^{ABS}=\sum_m\sum_\mu \Bigg[\Big[U_{ABS}^\dagger\Big(E_\mu^{AB}\otimes F_m^S\Big)U_{ABS}\Big]\rho_1^{ABS}\Big[U_{ABS}^\dagger\Big(E_\mu^{AB}\otimes F_m^S\Big)U_{ABS}\Big]^\dagger\Bigg].
		\end{align}
		Let's consider a $K_{\mu m}^{ABS}=U_{ABS}^\dagger\Big(E_\mu^{AB}\otimes F_m^S\Big)U_{ABS}$, which is the kraus operator applied on the initial state ($\rho_1^{ABS}$). Then, we can re-write the state $\rho_4^{ABS}$ as,
		\begin{align}
			\rho_4^{ABS}=\sum_{\mu m}K_{\mu m}^{ABS}\rho_1^{ABS}{K_{\mu m}^{ABS}}^\dagger.
		\end{align}
		The explicit form of the kraus operator $K_{\mu m}^{ABS}$ is,
		\begin{align}
			K^{ABS}_{\mu m}=U_{ABS}^\dagger\left(\sum_{j,k=0}^1q_{\mu jk} \left(\sigma_z^j\sigma_x^k\right)^A\otimes \left(\sigma_x^{2-j}\right)^B\otimes \sum_{i=0}^3 c_{mi} \sigma_i^S \right)U_{ABS}.
		\end{align}
		The explicit form of $K_{\mu m}^{ABS}$ is, 
		\begin{align*}
			K_{\mu m}^{ABS}=\Big[q_{\mu 00}\Big(\mathds{1}^A\otimes\mathds{1}^B\otimes c_{m0}\mathds{1}^S+\mathds{1}^A\otimes\sigma_z^B\otimes c_{m1}\sigma_x^S+\sigma_z^A\otimes\sigma_z^B\otimes c_{m2}\sigma_y^S+\sigma_z^A\otimes\mathds{1}^B\otimes c_{m3}\sigma_z^S\Big)+\\
			q_{\mu 01}\Big(\sigma_x^A\otimes\sigma_z^B\otimes c_{m0}\sigma_x^S+\sigma_x^A\otimes\mathds{1}^B\otimes c_{m1}\mathds{1}^S+\sigma_y^A\otimes\mathds{1}^B\otimes c_{m2}\sigma_z^S-\sigma_y^A\otimes\sigma_z^B\otimes c_{m3}\sigma_y^S\Big)+\\
			q_{\mu 10}\Big(\sigma_z^A\otimes\sigma_x^B\otimes c_{m0}\sigma_z^S+\sigma_z^A\otimes\sigma_y^B\otimes c_{m1}\sigma_y^S-\mathds{1}^A\otimes\sigma_y^B\otimes c_{m2}\sigma_x^S+\mathds{1}^A\otimes\sigma_x^B\otimes c_{m3}\mathds{1}^S\Big)+\\
			q_{\mu 11}\Big(\sigma_y^A\otimes\sigma_y^B\otimes c_{m0}\sigma_y^S+\sigma_y^A\otimes\sigma_x^B\otimes c_{m1}\sigma_z^S-\sigma_x^A\otimes\sigma_x^B\otimes c_{m2}\mathds{1}^S-\sigma_x^A\otimes\sigma_y^B\otimes c_{m3}\sigma_x^S\Big)
			\Big]
		\end{align*}
		Without loss of generality, we may consider the initial system state to be a pure state $\rho^S=\proj{\psi}^S$. Then the overall transformation becomes
		\begin{align}
			\rho_{ABS}^4=\sum_{\mu m}\Big(K_{\mu m}^{ABS}\Big)\proj{++}^{AB}\otimes\proj{\psi}^S\Big(K_{\mu m}^{ABS}\Big)^\dag
		\end{align}
		To simply understand, we write the effect of each Kraus operator on the global initial state as, 
		\begin{align*}
			K_{\mu m}^{ABS} \left(\ket{++}^{AB} \otimes \ket{\psi^S}\right)=\Big(q_{\mu00}c_{m0}+q_{\mu01}c_{m1}+q_{\mu10}c_{m3}-q_{\mu11}c_{m2}\Big)\ket{++}\otimes\mathds{1}\ket{\psi_s}+\\
			\Big(q_{\mu00}c_{m1}+q_{\mu01}c_{m0}+iq_{\mu10}c_{m2}+iq_{\mu11}c_{m3}\Big)\ket{+-}\otimes\sigma_x\ket{\psi_s}+\\
			\Big(q_{\mu00}c_{m3}-iq_{\mu01}c_{m2}+q_{\mu10}c_{m0}-iq_{\mu11}c_{m1}\Big)\ket{-+}\otimes\sigma_z\ket{\psi_s}+\\
			\Big(q_{\mu00}c_{m2}+iq_{\mu01}c_{m3}-iq_{\mu10}c_{m1}-q_{\mu11}c_{m0}\Big)\ket{--}\otimes\sigma_y\ket{\psi_s}.
		\end{align*} 
		
		\item[Step 5:] Finally the state $\rho_{ABS}^4$ is evolved with the acausal unitary $V_{ABS}$ given by
		\begin{align}
			V_{ABS}&=H_A\otimes H_B\otimes H_S \Big(U_{ABS}\Big)H_A^\dagger\otimes H_B^\dagger\otimes H_S^\dagger \\
			&=\proj{++}^{AB} \otimes \mathds{1}^S+ \proj{+-}^{AB} \otimes \sigma_x^S + \proj{-+}^{AB} \otimes \sigma_z^S+i \proj{--}^{AB} \otimes \sigma_y^S,
		\end{align}
		where $H_X$ with X= A,B,S is the Hadamard unitary. As a result, the final state becomes
		\begin{align}
			\rho_{ABS}^5=V_{ABS} \ \rho_{ABS}^4 \ V_{ABS}^\dag= \Psi^{AB}\Big(\ketbra{++}{++}^{AB}\Big) \otimes \rho^S,
		\end{align}
		where $\Psi^{AB}\Big(\ketbra{++}{++}^{AB}\Big)=\sum_\mu F^{AB}_{\mu m} \ketbra{++}{++}^{AB} F_{\mu m}^{AB \dag}$ is a channel implemented on the ancilla $AB$, with the Kraus operators
		\begin{align}
			F_{\mu m}^{AB}=q_{\mu00}\Big(c_{m0}\mathds{1}\otimes\mathds{1}+c_{m1}\mathds{1}\otimes\sigma_z+c_{m2}\sigma_z\otimes\sigma_z+c_{m3}\sigma_z\otimes\mathds{1}\Big) \nonumber \\+q_{\mu01}\Big(c_{m0}\sigma_x\otimes\sigma_z+c_{m1}\sigma_x\otimes\mathds{1}+c_{m2}\sigma_y\otimes\mathds{1}-c_{m3}\sigma_y\otimes\sigma_z\Big) \nonumber \\
			+q_{\mu10}\Big(c_{m0}\sigma_z\otimes\sigma_x+c_{m1}\sigma_z\otimes\sigma_y-c_{m2}\mathds{1}\otimes\sigma_y+c_{m3}\mathds{1}\otimes\sigma_x\Big) \nonumber \\+q_{\mu11}\Big(c_{m0}\sigma_y\otimes\sigma_y+c_{m1}\sigma_y\otimes\sigma_x-c_{m2}\sigma_x\otimes\sigma_x-c_{m3}\sigma_x\otimes\sigma_y\Big).
		\end{align}  
	\end{enumerate}

	Thus, the arbitrary system state is fully recovered, as if the system is passed through a transparent environment.  On the level of operation the transformation leads to, after all the steps of protocol.
	\begin{align}
		\Phi^{AB}\otimes\Lambda^S \rightarrow \Psi^{AB}\otimes\mathds{1}^S.
	\end{align}
	The same transparency can be attained by applying a non-unitary  (semi-causal) CPTP operation on $ABS$ with the Kraus operators  $\{ \proj{++}^{AB} \otimes \mathds{1}^S, \ \proj{+-}^{AB} \otimes \sigma_x^S, \ \proj{--}^{AB} \otimes \sigma_y^S,  \ \proj{-+}^{AB} \otimes \sigma_z^S\}$ in Step 5, instead of $V_{ABS}$.

\section{The protocol for qudit channels}
\label{app:Qudit calculation}
	
	{\bf The protocol for qudit channels ($d \geqslant 3$)} --
	The protocol for a $d$-dimensional system ($S$) follows similar to the qubit systems, where an arbitrary qudit channel $\Lambda_d^S$ is made transparent with the help of two $d$-dimensional ancillary systems  $A$ and $B$. In Step 1,  the ancilla state is prepared in $\ket{\psi_0}^A\otimes\ket{\psi_0}^B$ where  $\ket{\psi_0}=\frac{1}{\sqrt{d}}\sum_{k=0}^{d-1} \ket{k}$. In Step 2, the tri-partite unitary operation  $U^{(d)}_{ABS} = (\mathds{1}_B \otimes C^{AS}_X) (\mathds{1}_A \otimes C^{BS}_Z) $ is applied. Here the control operations $C^{AS}_X$ and $C^{BS}_Z$ are defined as
	\begin{align}
		C^{BS}_Z=\sum_{k=0}^{d-1} \proj{k}^B\otimes Z^k, \ \ 
		C^{AS}_X=\sum_{k=0}^{d-1} \proj{k}^A\otimes X^k,
	\end{align}
	where $Z=\sum_{k=0}^{d-1} e^{(i2k\pi/d)} \proj{k}$ and $X=\sum_{k=0}^{d-1}  \ket{(k+1) \ \mbox{mod} \ d} \bra{k}$. In Step 3, the system is exposed to an arbitrary environment and undergoes a noisy operation $\Lambda_d^S$. The ancilla $AB$ may also undergo noisy operation $\Phi_d^{AB}$ with the corresponding Kraus operators
	\begin{align}
		E^{AB}_\mu=\sum_{i,j}c_{\mu ij} Z^i X^j \otimes X^{d-i}. 
	\end{align}
	In Step 4, the $ABS$ composite is evolved with the unitary $U^{(d)\dag}_{ABS}$ and followed by  $V^{(d)}_{ABS}$, where 
	\begin{align}
		V^{(d)}_{ABS}= \sum_{m, n} \proj{\psi_m}^A \otimes \proj{\psi_n}^B \otimes (Z^m X^n)^\dagger,
	\end{align}
	with $\ket{\psi_m}^{A}=Z^m\ket{\psi_0}^{A}$ and $\ket{\psi_n}^B=Z^{[n(d-1) \mod  d]}\ket{\psi_0}^B$. The Steps 1-4 result in the overall transformation on the level of channel as
	\begin{align}
		\Phi_d^{AB} \otimes \Lambda_d^S \to \Psi_d^{AB} \otimes \mathds{1}^S,
	\end{align}
	where the local channel on the system $S$ becomes transparent (see Appendix for more details). \\
	
	Suppose, an arbitrary noisy quantum channel applied on a $d$-dimensional quantum system $S$ in the state $\eta^{S}$, given by 
	$\Lambda_d^{S}(\eta^{S})=\sum_{i} F_{i}(\eta^{S})F_{i}^\dagger$. Now, we extend the protocol to that makes the channel transparent, i.e., $\Lambda_d^{S} \to \mathds{1}^S$. Note, any Karus operator can be expressed as $F_{i}=\sum_{m,n} c_{mni} S_{mn}$, where $S_{mn}$ are the complete set of Schwinger unitary operators, given by $S_{mn}=Z^{m}X^{n}$ with $Z=\sum _{k=0}^{d-1} \xi^{k}\proj{k}$, $X=\sum_{k=0}^{d-1}\ket {(k+1)\hspace{0.1cm}mod\hspace{0.1cm} d} \bra{k}$. Here $\{\ket{k}\}$ represents a complete set orthonormal bases in the system Hilbert space.
	
	For the protocol, we attach two $d$-dimensional anilla $AB$ with the system $S$, and the ancilla allowed to interact with a wide class of environment. Similar to the qubit case, the protocol is implemented in five steps, as follows.

	\begin{enumerate}
		\item[Step 1:] Initially, we attach ancilla $AB$ with system state $S$.
		\begin{align}
			\eta_{1}^{ABS}= \proj{\psi_0}^{A}\otimes \proj{\psi_0}^{B}\otimes \eta^{s},	
		\end{align}
		where $\ket{\psi_0}_{A/B}=\frac{1}{\sqrt{d}}\sum_{k=0}^{d-1}\ket{k}_{A/B}$ and $\eta^S$ can be arbitrary state for the system. Here $\{\ket{k}_{A/B}\}$ is a complete set of orthonormal bases on the Hilbert space of $A/B$. 
		
		\item[Step 2:] The Initial state $\eta_{1}^{ABS}$ is evolved with global (acausal) unitary $U_{ABS}$, given by
		\begin{align}
			U_{ABS}=U_{AS}^{X}U_{BS}^{Z}=\sum_{kk'}^{d^2-1}\proj{kk'}\otimes X^kZ^{k'},	
		\end{align}
		where where $	U_{AS}^{X}=\sum_{k=0}^{d-1}\proj{k}^{B}\otimes X^{k}$, $U_{BS}^{Z}=\sum_{k'=0}^{d-1}\proj{k'}^{A}\otimes Z^{k'}$. As a consequence, the initial state is transformed to $\eta_{2}^{ABS}=U_{ABS}\eta_{1}^{ABS}U_{ABS}^\dagger$.
		
		\item[Step 3:] Now, the system $S$ is passed through an arbitrary noisy environment implementing a CPTP map  $\Lambda_d^S$. The ancilla may also undergo a wide class of noisy channel $\Phi_d^{AB}$ due to interaction with its environment. As a result,
		\begin{align}
			\eta_{3}^{ABS}= \Phi^{AB}_d\otimes \Lambda_d^{S}(\eta_{2}^{ABS}),
		\end{align}
		where the kraus operators of the maps $\Phi_d^{AB}$ and $\Lambda_d^S$ are $E_{\mu}^{AB}=\sum_{\alpha, \beta}p_{\mu\alpha\beta}Z^{\alpha}X^{\beta}\otimes X^{d-\alpha}$ and $F^S_{i}=\sum_{mn}c_{i m n }Z^m X^n$ respectively, with $p_{\mu\alpha\beta} \in \mathds{C}$ and $c_{imn}\in \mathds{C}$. Note, the Kraus operators satisfy the trace preserving condition $\sum_\mu {E^{AB}_\mu}^\dagger E_\mu^{AB}=\mathds{1}^{AB}$ and similarly $\sum_i {F^S_i}^\dagger F_i^S=\mathds{1}^S$.
		
		\item[Step 4:] The global unitary $U_{ABS}^\dagger$ is applied on  $\eta_{3}^{ABS}$, to give
		\begin{equation}
			\eta_{4}^{ABS}=U_{ABS}^\dagger \eta_{3}^{ABS} U_{ABS}. \label{eq:B5}
		\end{equation}
		Note, it is equivalent to write
		\begin{align}
			\eta_{4}^{ABS}=\sum_{\mu i}\Bigg[U_{ABS}^\dagger \Big(E_\mu^{AB}\otimes F_i^S\Big)U_{ABS}\Bigg]\eta_{1}^{ABS}\Bigg[U_{ABS}^\dagger \Big(E_\mu^{AB}\otimes F_i^S\Big)U_{ABS}\Bigg]^\dagger=\sum_{\mu i}K_{\mu i}^{ABS}\eta_{1}{K_{\mu i}^{ABS}}^\dagger
		\end{align}
		where Kraus operators $K_{\mu i}^{ABS}$ are applied on the initial state $\eta_{1}^{ABS}$. The Kraus operators are
		\begin{equation}
			K_{\mu i}^{ABS}=U_{ABS}^\dagger \left(\sum_{\alpha, \beta}^{d-1}p_{\mu\alpha\beta}\Big(Z^{\alpha}X^{\beta}\Big)^A\otimes \Big({X^{d-\alpha}}\Big)^{B} \otimes \sum_{m,n}^{d-1} c_{imn} S_{mn}\right)U_{ABS}
		\end{equation}
		One can simplify $K_{\mu i}^{ABS}$ by expressing the $U_{ABS}$ and $S_{mn}$ in terms of $Z$ and $X$. For simplicity, we consider that the ancilla goes through a identity channel ($\Phi^{AB}_d=\mathds{1}$). In that case, $K_{\mu i}^{ABS}\equiv K_i^{ABS}$. However, the protocol still works even after the ancilla still undergoes a wide class if noisy channels, as mentioned above. Now,
		\begin{align}
			K_{i}^{ABS}=\sum_{k,k'=0}^{d^2-1}\sum_{m,n=0}^{d-1} c_{mni}\proj{kk'}^{AB}\otimes {{Z}^{k'}}^\dagger {{X}^{k}}^\dagger Z^{m}X^{n}{X}^{k}{Z}^{k'}= \sum_{k,k'=0}^{d^2-1}\sum_{m,n=0}^{d-1}\left(c_{mni} \proj{kk'}^{AB}\otimes e^{i2\pi (km-k'n)/{d}}S_{m n}\right).
		\end{align}
		Here we have used the relations 
		\begin{align}
			Z^{k'}X^{k}=e^{{i2\pi kk'}/{d}} X^{k}Z^{k'}, \ \ 
			{Z^{k'}}^\dagger {X^{k}}^\dagger =e^{{i2\pi kk'}/{d}} {X^{k}}^\dagger {Z^{k'}}^\dagger, \ \
			{X^{\alpha}}^{\dagger}Z^{\beta}=e^{{i2\pi \alpha \beta }/{d}}Z^{\beta}{X^{\alpha}}^{\dagger}, \ \
			{Z^{\beta}}^{\dagger}X^{\alpha}=e^{{-i2\pi \alpha \beta }/{d}}X^{\alpha}{Z^{\beta}}^{\dagger}.
		\end{align}
		Upon rearrangements, the $K^{ABS}_i$ further reduces to
		\begin{align}
			K_i^{ABS}= \sum_{m,n=0}^{d-1}c_{mni}\Big[{Z^m}\otimes {Z^{n(d-1) mod\hspace{0.1cm}d}}\otimes{S_{m,n}}\Big].
		\end{align} 
		
		Now to understand the action of these Kraus operators, we assume that the system $S$ is initially in an arbitrary pure state $\ket{\phi}^S$. Note, any  mixed state can purified in a larger Hilbert space. Then,
		\begin{align}
			\ket{\Psi_4}^{ABS}=K_i^{ABS} \Big(\ket{\psi_{0}}^A\otimes\ket{\psi_{0}}^B\otimes\ket{\phi}^S\Big) \nonumber =\sum_{m,n=0}^{d-1}c_{imn}\ket{\psi_{m}}^A\otimes\ket{\psi_{n}}^B\otimes S_{mn}\ket{\psi}^S, \label{eq:B17}
		\end{align}
		where $\braket{\psi_{i}}{\psi_{j}}_{A/B}=\delta_{ij}$ and  $\ket{\psi_{m}}^A=Z^m\ket{\psi_{0}}^A$ and $\ket{\psi_{n}}^B={Z^{n(d-1) mod\hspace{0.1cm}d}}\ket{\psi_{0}}^B$. 
		
		\item[Step 5:] In the last step, we apply a global unitary $V^{(d)}_{ABS}$, given by
		\begin{align}
			V_{ABS}^{(d)}=\sum_{m,n}\proj{\psi_{m}}\otimes\proj{\psi_{n}}\otimes S_{mn}^\dagger,
		\end{align}
		and, as a result, the overall transformation becomes
		\begin{align}
			V^{(d)}_{ABS} \ \eta_4^{ABS} \ V^{(d) \dag}_{ABS} = \Psi_d^{AB} \Big(\proj{\psi_0}^A \otimes \proj{\psi_0}^B \Big) \otimes \eta^S,
		\end{align}
		where the Kraus operators corresponding to the CPTP map $\Psi_d^{AB}$ can easily found.
		
		Thus on the level quantum channels, the protocol implements the transformation
		\begin{align}
			\Phi_d^{AB} \otimes \Lambda_d^S \rightarrow \Psi_d^{AB}\otimes \mathds{1},
		\end{align}
		for arbitrary quantum channel of system $S$ and a wide class of noisy channel on the ancilla $AB$.
	\end{enumerate}

\end{document}